\RequirePackage{fixltx2e}
\documentclass[twocolumn,superscriptaddress]{revtex4-1}
\usepackage{amsmath}    
\usepackage{bm}
\usepackage{graphicx}   
\usepackage{verbatim}   
\usepackage{color}      
\usepackage{subfigure}  
\usepackage{hyperref}   
\begin{document}

\title{Mapping Skyrmion Stability in Uniaxial Lacunar Spinel Magnets from First--Principles}

\author{Daniil A. Kitchaev}
\email{dkitch@ucsb.edu}
\affiliation{Materials Department and Materials Research Laboratory, University of California, Santa Barbara, California 93106, USA}

\author{Emily C. Schueller}
\affiliation{Materials Department and Materials Research Laboratory, University of California, Santa Barbara, California 93106, USA}

\author{Anton Van der Ven}
\email{avdv@ucsb.edu}
\affiliation{Materials Department and Materials Research Laboratory, University of California, Santa Barbara, California 93106, USA}

\begin{abstract}
The identification of general principles for stabilizing magnetic skyrmion phases in bulk materials over wide ranges of temperatures is a prerequisite to the development of skyrmion--based spintronic devices. Lacunar spinels with the formula GaM${}_4$X${}_8$ with M=V, Mo; X=S, Se are a convenient case study towards this goal as they are some of the first bulk systems suggested to host equilibrium chiral skyrmions far from the paramagnetic transition. We derive the magnetic phase diagrams likely to be observed in these materials, accounting for all possible magnetic interactions, and prove that skyrmion stability in the lacunar spinels is a general consequence of their crystal symmetry rather than the details of the material chemistry. Our results are consistent with all experimental reports in this space and demonstrate that the differences in the phase diagrams of particular spinel chemistries are determined by magnetocrystalline anisotropy, up to a normalization factor. We conclude that skyrmion formation over wide ranges of temperatures can be expected in all lacunar spinels, as well as in a wide range of uniaxial systems with low magnetocrystalline anisotropy.
\end{abstract}

\maketitle


The prediction and experimental demonstration of topologically non--trivial magnetic structures, commonly called magnetic skyrmions\cite{Bogdanov1989, Roessler2006, Muhlbauer2009}, has sparked considerable interest in the conditions required for the formation of these phases, and their potential for applications in spintronic devices\cite{Jonietz12010, Sampaio2013}. Most reports of skyrmion formation in bulk systems have focused on cubic helimagnets, most frequently with the B20 structure: MnSi\cite{Muhlbauer2009}, FeGe\cite{Yu2012}, Cu${}_2$OSeO${}_3$\cite{Seki2012} and Co${}_x$Zn${}_y$Mn${}_z$\cite{Tokunaga2015}, where skyrmions are typically observed in a narrow range of fields and temperatures near the paramagnetic transition. Much larger stability windows for skyrmions have been reported in thin--film systems\cite{Yu2010}, and recently in the bulk uniaxial helimagnet GaV${}_4$Se${}_8$\cite{Fujima2017, Bordacs2017} and Heusler alloys such as Mn${}_{1.4}$PtSn\cite{Nayak2017}. The realization of practical skyrmion--based spintronic devices requires the wide stability windows observed in these materials, motivating a search for general principles leading to formation of thermally--robust skyrmion phases.

The lacunar spinels GaM${}_4$X${}_8$ are a convenient model system for studying mechanisms leading to robust skyrmion stability at all temperatures below the Curie temperature. GaV${}_4$S${}_8$\cite{Kezsmarki2015} and GaV${}_4$Se${}_8$\cite{Fujima2017, Bordacs2017} have been reported as skyrmion hosts with unusually wide thermal stability windows, while GaMo${}_4$S${}_8$ has been suggested as a skyrmion host on the basis of computational data\cite{Zhang2019}. Materials in this class exhibit significant metal--metal bonding\cite{Schueller2019, Streltsov10491}, with electronic structure defined by isolated M${}_4$ molecular units. Their magnetic behavior is well described by interactions between effective spins centered on the M${}_4$ clusters\cite{Kim2014}. Furthermore, the R3m symmetry and strong Dzyaloshinskii--Moriya interactions (DMI) common to these systems guarantee that skyrmion formation can be treated with explicit spin models as a largely two--dimensional problem.

Here, we demonstrate that skyrmion stability in GaM${}_4$X${}_8$ is a consequence of the symmetry of these materials rather than specifics of the magnetic interactions, with the exception of magnetocrystalline anisotropy. We construct a field--temperature magnetic phase diagram for the lacunar spinels based on a general cluster expansion Hamiltonian parametrized using density--functional theory (DFT) data\cite{Sanchez1984, vandeWalle2002, Drautz2004, Mueller2006, Janson2016, VdV2018}, which we find to be in close agreement with experimental reports. By analyzing the sensitivity of the phase diagram to all symmetrically--allowed perturbations in the Hamiltonian, we find that the form of the phase diagram is largely controlled by uniaxial anisotropy, as well as higher--order in--plane anisotropy. In the low--anisotropy regime, skyrmion formation is guaranteed by the lack of a competing canted spin--wave phase magnetized along the high--symmetry axis, consistent with phenomenological predictions\cite{Bogdanov1994}, which leads us to conclude that the phase behavior we compute is likely to be broadly applicable to uniaxial magnets with strong in--plane DMI.

\section*{Methods}
\subsection*{Cluster expansion generation and fitting}
We construct the magnetic cluster expansion following a methodology similar to that described by Thomas and Van der Ven\cite{Thomas2017, VdV2018, Thomas2018}. Our cluster expansion of the internal energy only includes magnetic degrees of freedom, where the moment on each M${}_4$ tetrahedron is represented by a 3--dimensional unit vector. To construct the cluster expansion, we identify all site--clusters up to a target radius and number of sites, and the symmetry operations which map each cluster to itself. Then, we generate all possible basis functions for spin interactions on each cluster. Following previous derivations of cluster expansions for orientational degrees of freedom\cite{Drautz2004, Mueller2006, Singer2011}, we use products of spherical harmonics $|l,m\rangle = \sqrt{4\pi}Y^{l}_{m}(\phi, \theta)$ as a complete basis set for spin interactions, where $(\phi, \theta)$ are the spin vector orientation in spherical coordinates. We additionally group these products according to the total symmetry of the interaction to form basis functions of the form:
\[
|l_1,l_2;L,M\rangle = 4\pi \sum_{m_1,m_2} c^{l_1,l_2,L}_{m_1,m_2,M}Y^{l_1}_{m_1}(\phi_1, \theta_1)Y^{l_2}_{m_2}(\phi_2, \theta_2)
\]
where $c^{l_1,l_2,L}_{m_1,m_2,M}$ are Clebsch--Gordan coefficients. This procedure isolates the basis functions corresponding to exchange ($L=0$), DMI ($L>0$, odd), and anisotropy ($L>0$, even). Finally, we find the purely real component of each basis function invariant to the symmetry of the cluster, and using Gram-Schmidt orthogonalization, obtain an orthonormal basis set for spin interactions on each cluster. The cluster expansion implementation relies on an in--house python code accelerated using the Numba package\cite{Numba2015}, while general structure processing, data handling, and symmetry analysis rely on the pymatgen package\cite{Ong2013}. A detailed description of the cluster expansion and the procedure used to generate interaction functions is available in Supplementary Note 1.\cite{SupplementaryMat}

To obtain a fit for the interaction coefficients in the cluster expansion from  DFT data, we follow a standard methodology designed for the automated generation of phase diagrams\cite{vandeWalle2002, Hart2005}. First, we enumerate symmetrically distinct collinear and spin--wave configurations compatible with supercells up to size 4. We refine this dataset by identifying the least--constrained correlation vectors in the input data as the eigenvectors of the correlation covariance matrix with the smallest eigenvalues. We then obtain spin--configurations corresponding to these correlation vectors and add them to the fitting dataset. Finally, we fit the cluster expansion interaction coefficients using least--squares regression, while using a genetic algorithm to eliminate basis functions from the Hamiltonian so as to maximize the cross--validation (CV) score.

\subsection*{Monte Carlo sampling and ground state search}
We use a Hamiltonian Monte Carlo approach to sample the finite--temperature behavior given by the cluster expansion Hamiltonian. Our Monte Carlo implementation exactly follows the formalism described by Wang \emph{et al}\cite{Drautz2019}, with trajectory sampling based on the No U-Turn Sampler/Dynamic Multinomial Sampling methods\cite{Hoffman2014, Betancourt2017}. All Monte Carlo runs reported here are constant--field heating runs, where each temperature step first rejects 800 uncorrelated samples for equilibration, and then saves 2000 uncorrelated samples for production. To ensure that the obtained samples are uncorrelated, we set the number of Monte Carlo passes between samples to exceed the estimated autocorrelation decay time.

To identify ground state spin configurations, we first generate candidate structures using simulated annealing starting from a random configuration and representative configurations of known phases. We then relax each configuration to its local minimum using conjugate gradient minimization, and save the lowest energy structure.

\subsection*{Density functional theory calculations}
DFT calculations are performed using the Vienna Ab-Initio Simulation Package (VASP) \cite{Kresse1996a}, using the projector-augmented-wave method \cite{Kresse1999} with the Perdew-Burke-Ernzerhof (PBE) exchange--correlation functional \cite{PBE1996}. We do not apply a Hubbard--$U$ correction because our previous benchmarks on GaV${}_4$Se${}_8$ revealed that the standard on--site Hubbard--$U$ approach leads to an incorrect electronic configuration and magnetic behavior\cite{Schueller2019}. All calculations account for spin--orbit coupling and are converged to 10${}^{-6}$ eV in total energy. We use a reciprocal space discretization of 100 k-points per \AA${}^{-3}$, and smearing width of 0.05 eV based on a convergence of total energy across all distinct supercells containing 2 formula units of GaMo${}_4$S${}_8$ to 0.5 meV/f.u. To further reduce error arising from changes in the k-point mesh across different supercells, we reference all magnetic configuration energies to that of a $c$--axis ferromagnet computed using the same supercell. In all cases, DFT calculations are done statically, based on the experimentally--observed low--temperature structure.

\section*{Results}

\subsection*{Magnetic cluster expansion Hamiltonian}

\begin{figure*}[t]
\includegraphics[width=\textwidth]{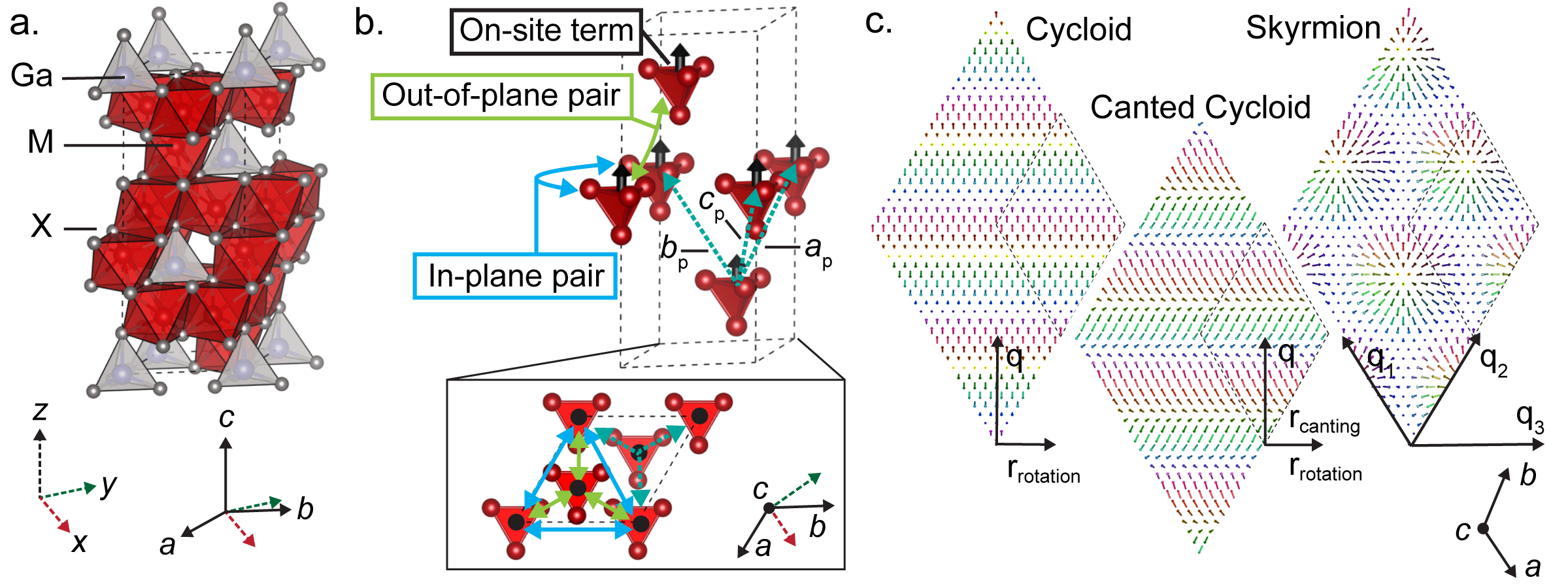}
\caption{\label{fig:structure} \textbf{a.} Low-temperature structure of a GaM${}_4$X${}_8$ lacunar spinel in the R3m conventional unit cell. \textbf{b.} Model for the magnetic structure, treating M${}_4$ tetrahedra as magnetic units forming a face--centered--cubic lattice, where the magnetic Hamiltonian consists of by single-site and nearest-neighbor interaction energies. Lattice vectors ($a_{\text{p}}$, $b_{\text{p}}$, $c_{\text{p}}$) shown form the primitive lattice used to define the cluster expansion Hamiltonian. \textbf{c.} Example configurations of helimagnetic phases observed in lacunar spinels, where $q$ denotes a propagation wavevector and color corresponds to spin orientations.}
\end{figure*}

\begin{table}
\begin{ruledtabular}
\begin{tabular}{l r l c}
Cluster type & \multicolumn{2}{c}{Basis function}  & $J^{(0)}$(meV) \\ \hline
On-site		& $\phi^{A}_1=$	& $\sqrt{2} |2,0\rangle$ & 0 \\
$r = (0, 0, 0)$	& $\phi_2^{A}=$		& $-i \left(|4,-3\rangle + |4,3\rangle\right) $	& 0.03(5)\\
(1 equiv.)			& $\phi^{A}_3=$		& $\sqrt{2} |4,0\rangle$ & 0 \\
			& $\phi^{A}_4=$		& $|6,-6\rangle + |6,6\rangle$	& 0 \\
			& $\phi^{A}_5=$		& $-i \left(|6,-3\rangle + |6,3\rangle\right)$ & 0 \\
			& $\phi^{A}_6=$		& $\sqrt{2}|6,0\rangle$	& 0 \\ \hline
Out-of-plane& $\phi^{E}_7=$		& $\frac{\sqrt{2}}{3} |1,1;0,0\rangle$	& 0.62(5) \\
$r_1 = (0, 0, 0)$	& $\phi^{D}_8=$		& $\frac{i}{3} \left( |1,1;1,1\rangle - |1,1;1,-1\rangle \right)$ & 0.88(9) \\
$r_2 = (-1, 0, 0)$	& $\phi^{A}_9=$		& $-\frac{i}{3} \left( |1,1;2,1\rangle + |1,1;2,-1\rangle \right)$ & 0  \\
(3 equiv.)			& $\phi^{A}_{10}=$	& $\frac{\sqrt{2}}{3} |1,1;2,0\rangle $ & 0 \\
			& $\phi^{A}_{11}=$	& $\frac{1}{3} \left( |1,1;2,2\rangle + |1,1;2,-2\rangle \right)$ & 0 \\  \hline
In-plane	& $\phi^{E}_{12}=$	& $\frac{\sqrt{2}}{3} |1,1;0,0\rangle$ & 1.07(5) \\
$r_1 = (0,0,0)$	& $\phi^{D}_{13}=$	& $(\frac{1}{6} + \frac{i}{\sqrt{12}})|1,1;1,1\rangle +	$ & 0.20(9) \\
$r_2 = (-1,0,1)$		&				& $(\frac{1}{6} - \frac{i}{\sqrt{12}})|1,1;1,-1\rangle$ &  \\
(3 equiv.)			& $\phi^{D}_{14}=$	& $\frac{-i\sqrt{2}}{3} |1,1;1,0\rangle$ & 0.08(7) \\
			& $\phi^{A}_{15}=$	& $\frac{\sqrt{2}}{3} |1,1;2,0\rangle$ & 0 \\
			& $\phi^{A}_{16}=$	& $(\frac{1}{\sqrt{12}}-\frac{i}{6})|1,1;2,1\rangle - $ & 0 \\
			& 				& $(\frac{1}{\sqrt{12}}+\frac{i}{6})|1,1;2,-1\rangle$ & \\
			& $\phi^{A}_{17}=$	& $(\frac{1}{6} - \frac{i}{\sqrt{12}})|1,1;2,2\rangle +	$ & 0.1(1) \\
			&				& $(\frac{1}{6} + \frac{i}{\sqrt{12}})|1,1;2,-2\rangle$ & \\

\end{tabular}
\end{ruledtabular}
\caption{\label{table:hamiltonian} Clusters and symmetrized basis functions for the GaM${}_4$X${}_8$ magnetic cluster expansion Hamiltonian, and the $J^{(0)}$ vector fitted to GaMo${}_4$S${}_8$ DFT data. Cluster site coordinates and basis functions are given for the reference cluster, in lattice coordinates with respect to the primitive lattice vectors ($a_{\text{p}}$, $b_{\text{p}}$, $c_{\text{p}}$) given in Fig \ref{fig:structure}b. The number of equivalents for each cluster refers to the number of symmetrically--equivalent clusters of this type per primitive cell. Basis functions are defined in terms of spherical harmonics $|l,m \rangle$ for the on-site terms and Clebsch--Gordan functions $|l_1,l_2;L,M \rangle$ for pair clusters $(r_1,r_2)$, as described in the methods. The cartesian form of the basis functions is available in Supplementary Table 1.\cite{SupplementaryMat} Basis function superscripts denote whether the interaction corresponds to exchange (E), DMI (D), or anisotropy (A). Parenthesis in the $J^{(0)}$ vector components denote uncertainty in the last digit.}
\end{table}

We begin by defining an effective spin Hamiltonian for the magnetic behavior of a GaM${}_4$X${}_8$ lacunar spinel in the form of a cluster expansion, which is a summation over interaction correlation functions $\varphi$ with interaction coefficients $J$. The correlation functions $\varphi$ are determined by the lattice type and symmetry of the material, while the interaction coefficients $J$ are specific to each chemistry. Thus, we can systematically explore the magnetic behavior of GaM${}_4$X${}_8$ by establishing which magnetic phase diagrams are likely to arise given the overall form of the Hamiltonian, across possible choices of interaction parameters $J$.

The full form of a cluster expansion Hamiltonian is
\[
E = \sum_{\Omega} \sum_{\alpha} J_{\alpha}^{\Omega} \sum_{\omega \in \Omega} \hat{p}_{\omega}\left[\phi_{\alpha}^{\Omega}\right] = \sum_{i} J_{i} \varphi_{i}
\]
where $\phi$ are interaction basis functions and $J$ are interaction coefficients. Each interaction is defined with respect to a cluster of sites $\omega$, where symmetrically--equivalent clusters are grouped into orbits $\Omega$. The interaction basis functions contain all spin--couplings consistent with the symmetry of the cluster, which include conventional Heisenberg exchange, DMI, and anisotropy interactions, as well as any higher--order terms. The symmetry operation $\hat{p}_{\omega}$ generates the cluster $\omega$ from a reference cluster for its orbit $\Omega$. The total contribution of a basis function $\phi$ for the symmetrically-equivalent clusters in $\Omega$ defines the correlation function $\varphi$.

We take the symmetry of the crystal to be R3m as shown in Figure \ref{fig:structure}a, which results from a low-temperature distortion of the $F\bar{4}3m$ vacancy--ordered spinel structure along the $\langle 111 \rangle$ direction. The magnetic sublattice consists of a distorted face--centered--cubic (FCC) arrangement of M${}_4$ tetrahedral clusters, shown in Figure \ref{fig:structure}b, where each M${}_4$ tetrahedron can be treated as a single spin vector. As the distance between M${}_4$ clusters is large, we approximate the magnetic energy with only on-site and nearest-neighbor couplings. For the three symmetrically--distinct couplings present (on-site, out-of-plane pair and in-plane pair), we derive spin--interaction basis functions consistent with the symmetry of each cluster. The basis functions are polynomials of the spin--vector components, up to sixth order for the on-site term, and bilinear order for the pair terms. These interactions, listed in Table \ref{table:hamiltonian}, form a complete basis set for the magnetic Hamiltonian that is applicable to any material with a relatively sparse FCC magnetic sublattice and R3m symmetry.

The coefficients $J$ of each correlation function $\varphi$ parametrize the variation of the Hamiltonian across different chemistries. Thus, in order to understand the phase behavior of all GaM${}_4$X${}_8$ lacunar spinels with R3m symmetry, it is sufficient to evaluate how the field--temperature phase diagram evolves with the components of $J$. We limit ourselves to $J$--vectors appropriate for locally ferromagnetic materials ($J_{7} > 0$, $J_{12} > 0$), allowing for strong spin-orbit coupling. This regime is characteristic of the behavior of skyrmion--hosting lacunar spinels with M=V,Mo and X=S,Se, where the observed magnetic phases are ferromagnet, cycloid, canted cycloid, and skyrmion. Example configurations of these phases are shown in Figure \ref{fig:structure}c.

\subsection*{Derivation of an example phase diagram for GaM${}_4$X${}_8$}

\begin{figure*}[t]
\includegraphics[width=\textwidth]{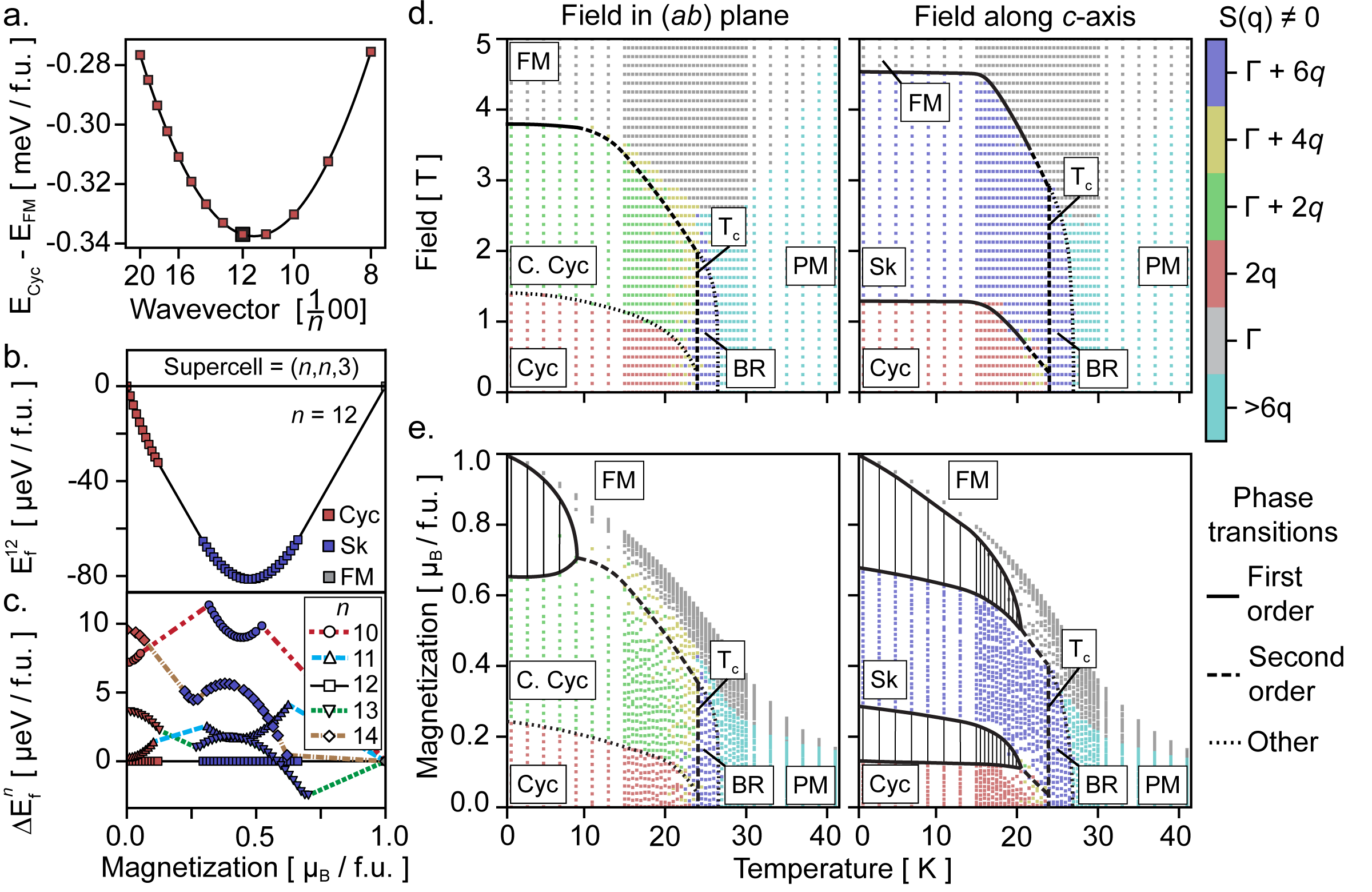}
\caption{\label{fig:gms} \textbf{a.} Energy of an ideal cycloid relative to that of a ferromagnet magnetized along the $c$-axis, as a function of $n$, which defines the cycloid propagation wavevector $q_n$ = [$\frac{1}{n}$00]. \textbf{b.},\textbf{c.} Formation energies of ground state configurations as a function of average magnetization along the $c$-axis, constrained to a ($n$,$n$,3) periodic supercell of the R3m unit cell shown in Fig. \ref{fig:structure}a. Formation energies are given with respect to the $n=12$ cycloid and $c$-axis ferromagnet (b.), and the full $n=12$ ground-state energy profile (c.). \textbf{d.},\textbf{e.} Magnetic phase diagram arising from the $J^{(0)}$ parametrization of the cluster expansion Hamiltonian, as a function of magnetic field magnitude (d.) and total magnetization (e.). Color denotes the number of $q$-points for which the structure factor $S(q)$ is non-zero. Phase labels correspond to ferromagnet (FM), cycloid (Cyc), canted cycloid (C. Cyc), skyrmion (Sk), Brazovskii region (BR) and paramagnet (PM). The locations and orders of phase boundaries are drawn based on Monte Carlo data to best agree with changes in $S(q)$, the topological index (see Supplementary Figure 1 \cite{SupplementaryMat}), discontinuities in internal energy and magnetization, and peaks in fluctuation data.  $T_\text{c}$ denotes the Curie temperature. Note that ``other'' denotes a change in structure factor $S(q)$ not accompanied by any discernible discontinuities in free energy.}
\end{figure*}

Our strategy for exploring phase behavior in this system is to construct a full field--temperature phase diagram for one choice of $J$--vector, and then calculate how perturbations in $J$ translate to changes in phase transitions. While this approach is only strictly valid for small deviations from the initial choice of $J$, the degree to which extrapolation is valid is determined by whether the form of the phase diagram is more determined by the values of $J$, or by which correlation functions $\varphi$ are present in the Hamiltonian. In this case, we will argue that given an appropriate normalization, the form of the correlation functions $\varphi$ plays the more important role, leading to a universal behavior of the phase diagram.

We choose our initial $J$--vector, $J^{(0)}$, by fitting the Hamiltonian to reproduce the magnetic behavior of GaMo${}_4$S${}_8$ as computed from DFT. We choose GaMo${}_4$S${}_8$ as a convenient reference point as a recent report has suggested that this material exhibits cycloid and skyrmion phases with particularly short wavelengths\cite{Zhang2019}, which allows us to directly study these phases with periodic-cell Monte Carlo. Furthermore, the electronic structure of GaMo${}_4$S${}_8$ appears to be reasonably captured by the standard PBE functional, while the better-known V-based analogs require more sophisticated, computationally expensive methods such as RPA\cite{Schueller2019}. Following a state--of--the--art cluster expansion fitting procedure, as well as a DFT calculation scheme designed to minimize spurious sources of error (details  available in the methods), we obtain the $J^{(0)}$ vector given in Table \ref{table:hamiltonian} with a RMSE of 0.3 meV/formula unit (f.u.) across a total energy range of 7 meV/f.u. The low absolute value of the error justifies our choice of truncating the Hamiltonian at nearest-neighbor interactions and bilinear pair couplings, as the inclusion of any additional basis functions would likely only be capturing noise in the DFT data. While both the total error and the uncertainty on the components of $J^{(0)}$ are small, the significance of these error bars in relation to phase behavior is not immediately clear. However, as our objective is to obtain a reasonable initial $J^{(0)}$ for our perturbative analysis, we proceed to characterize the phase diagram given by this fit and address the role of uncertainty, as well as general perturbations to $J$, in a later section.

We first establish the ground states of the $J^{(0)}$ Hamiltonian as a function of total magnetization, which include cycloid, canted cycloid, skyrmion and ferromagnet phases. The dominant periodicity of a helimagnet is set by the competition between DMI and exchange, which typically remains close to the period of the cycloid phase. The lowest energy commensurate cycloid in this system has wavevector $q = [\frac{1}{n} 0 0]$ for $n=12$, as shown in Figure \ref{fig:gms}a. We thereby choose a $(12,12,3)$ supercell of the conventional unit cell (1296 M${}_4$ units) for a full ground state enumeration, as this supercell is compatible with all low--energy cycloid variants, as well as the typical 6--fold skyrmion lattice phase. We obtain the internal energy profile shown in Figure \ref{fig:gms}b, which as a function of magnetization along the $c$-axis proceeds through the cycloid phase at low magnetization, skyrmion phase at intermediate magnetization, and ferromagnet phase at high magnetization. Repeating the ground state search for other $(n,n,3)$ supercells, we find that while the commensurate cycloid and canted cycloid phases are always minimized for $n=12$, the skyrmion phase relaxes from $n=12$ to $n=13$ at high magnetization ($\approx 8 \%$ change in wavelength), as shown in Figure \ref{fig:gms}c. This result is intriguing from the perspective of experimentally detecting skyrmions in diffraction data by means of a shift in magnetic structure $q$--vector away from that of a cycloid, and is consistent with observed changes in $q$ vector in the skyrmion phase of GaV${}_4$S${}_8$\cite{Kezsmarki2015}. However, for the purposes of thermodynamic stability calculations, the difference in energy between the $n=12$ and $n=13$ skyrmion is small enough to be negligible.

The finite--temperature phase diagram of the $J^{(0)}$ Hamiltonian is shown in Figure \ref{fig:gms}d,e as a function of applied field and observed magnetization respectively. The locations and orders of phase transitions are estimated based on changes in the magnetic structure factor, topological index, discontinuities in internal energy and magnetization, and peaks in fluctuation data. 

The low--field phase up to the Curie temperature ($T_\text{c}$) is a cycloid. Magnetization in the $(ab)$--plane leads to a continuous transition into a canted cycloid phase, followed by a transition to a ferromagnet. At low temperatures (approximately $T < 0.5 T_\text{c}$), the transition from canted cycloid to ferromagnet is first--order, while at higher temperatures this transition becomes second--order. We conclude that the order of the phase transition changes because we observe a peak in the magnetic susceptibility at this point at all temperatures, but the discontinuity in magnetization only exists below $0.5 T_\text{c}$. Magnetization along the $c$--axis leads to the formation of the skyrmion phase, followed by the ferromagnet phase. The formation of the topologically--nontrivial skyrmion phase is also confirmed by a change in the topological index from 0 to -1 as the $c$--axis magnetization is increased (see Supplementary Figure 1\cite{SupplementaryMat}). Both transitions are first-order at most temperatures, becoming second--order only close to $T_\text{c}$ (approximately $T > 0.8 T_\text{c}$). At low temperatures, skyrmions are stabilized with respect to cycloids enthalpically, consistent with the behavior of the ground--state configurations. However, above approximately $0.67 T_\text{c}$ the skyrmion region expands at the expense of the cycloid region indicating that at elevated temperatures, skyrmions are additionally stabilized entropically. Immediately above $T_\text{c}$, the cycloid, canted cycloid and skyrmion phase regions extend into a partially--disordered phase dominated by fluctuations in the ($ab$)--plane at the cycloidal $q$--vectors, as a two--dimensional analog of the Brazovskii region described in cubic helimagnets\cite{Brazovskii1975, Janoschek2013}. Note that despite having a similar structure factor to the skyrmion phase, the two--dimensional Brazovskii region is topologically trivial as can be seen in Supplementary Figure 1.\cite{SupplementaryMat} Finally, at higher temperatures, the system fully disorders to form a paramagnet.

\subsection*{Variation in phase stability with changes in the Hamiltonian}

\begin{figure*}[t]
\includegraphics[width=\textwidth]{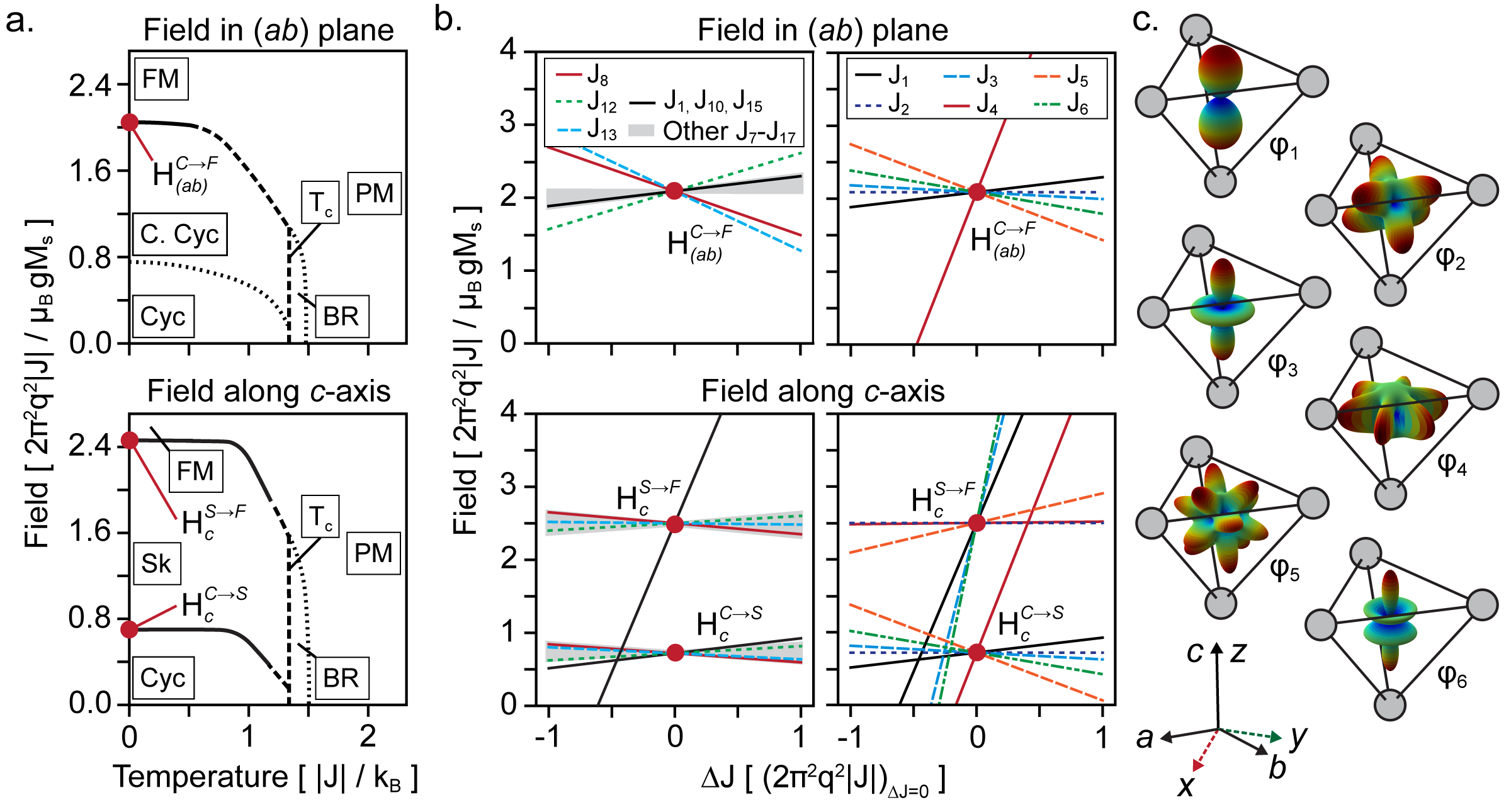}
\caption{\label{fig:cc} \textbf{a.} Normalized magnetic phase diagram derived using the $J^{(0)}$ Hamiltonian, marking three first-order phase transitions: cycloid--to--ferromagnet in the ($ab$) plane ($H^{ C\rightarrow F}_{(ab)}$), and cycloid--to--skyrmion and skyrmion--to--ferromagnet along the $c$ axis ($H^{ C\rightarrow S}_c$ and $H^{ S\rightarrow F}_c$ respectively). Phase labels are defined in the caption to Figure \ref{fig:gms}. \textbf{b.} Change in the normalized phase transition fields $H^{ C\rightarrow F}_{(ab)}$, $H^{ C\rightarrow S}_c$ and $H^{ S\rightarrow F}_c$ upon variation of the Hamiltonian parameters $J$. Note that the field normalization factor $2 \pi^2 q^2 |J| /\mu_{\text{B}} g M_s$ varies with $\Delta J$, while the units of $\Delta J$ [ $(2 \pi^2 q^2 |J|)_{\Delta J = 0 }$ ] are taken to be constant. \textbf{c.} Magnetocrystalline anisotropy functions $\varphi_{1,...,6}$ shown relative to a M${}_4$ tetrahedron, where the color and distance from the tetrahedron center along a certain direction represent the value of $\varphi_i$ for that  spin orientation.}
\end{figure*}

Having established the finite temperature phase diagram for one parametrization of the Hamiltonian, $J^{(0)}$, we evaluate how the phase diagram may change as the $J$ coefficients are varied. We first identify normalization factors for the phase diagram to account for changes in the Hamiltonian that amount to rescaling the field and temperature axes. We then use generalized Clausius--Clapeyron relationships to identify which components of the $J$ vector may alter the locations of first--order transitions seen in the phase diagram.

Figure \ref{fig:cc}a shows the phase diagram obtained for $J^{(0)}$, in units of the characteristic temperature and field for this system. The red dots highlight the three first--order transitions that define the low--temperature region of the phase diagram. The normalization factor for temperature is $|J| / k_{\text{B}}$, as any homogeneous rescaling of the Hamiltonian must also rescale temperature. The normalization factor for field is given by the characteristic difference in energy between the low-field and high-field ground states, which are the cycloid and ferromagnet phases respectively. In units of magnetic field, this factor is $2 \pi^2 q^2 |J| / \mu_{\text{B}} g M_s$, where $q$ is the magnitude of the cycloid wavevector in lattice coordinates and $M_s$ is the magnetic moment per spin. A full derivation is available in Supplementary Note 2.\cite{SupplementaryMat} Note that this factor is identical to the $D^2/A$ normalization used in previous literature where $D$ and $A$ are the effective DMI and exchange constants respectively.

Figure \ref{fig:cc}b plots how the locations of the three low--temperature first--order transitions change with variation in the components of $J^{(0)}$. The left panels account for the correlation functions corresponding to conventional bilinear spin couplings, while the right panels illustrate the effect of higher order on--site anisotropy terms, plotted schematically in Figure \ref{fig:cc}c. Note that $\varphi_1$ corresponds to quadratic single--spin anisotropy, while $\varphi_{10}$ and $\varphi_{15}$ are equivalent to XXZ anisotropy for out--of--plane and in--plane exchange respectively. In this case, these terms yield exactly the same behavior and are plotted as a single line. The change in the phase boundary location $H^\dagger / \frac{2 \pi^2 q^2 |J|}{\mu_B g M_s}$ is given by the generalized Clausius--Clapeyron relation,
\[
\frac{2\pi^2 q^2 |J|}{\mu_B g M_s} \frac{\partial \left(H^\dagger / \frac{2 \pi^2 q^2 |J|}{\mu_B g M_s}\right)}{\partial J} = \frac{\Delta \langle \varphi \rangle}{\Delta M} - \frac{2 H^\dagger}{q}\frac{\partial q}{\partial J} - H^\dagger J
\]
where $\Delta \langle \varphi \rangle$ and $\Delta M$ are the change in the correlation functions and magnetization across the first--order phase transition. Note that this expression explicitly accounts for the variation in $q$ and $|J|$ for the purposes of normalizing $H^\dagger$, while the change in $J$ is expressed in units of $(2\pi^2 q^2 |J|)_{\Delta J = 0}$.

\begin{figure*}[t]
\includegraphics[width=\textwidth]{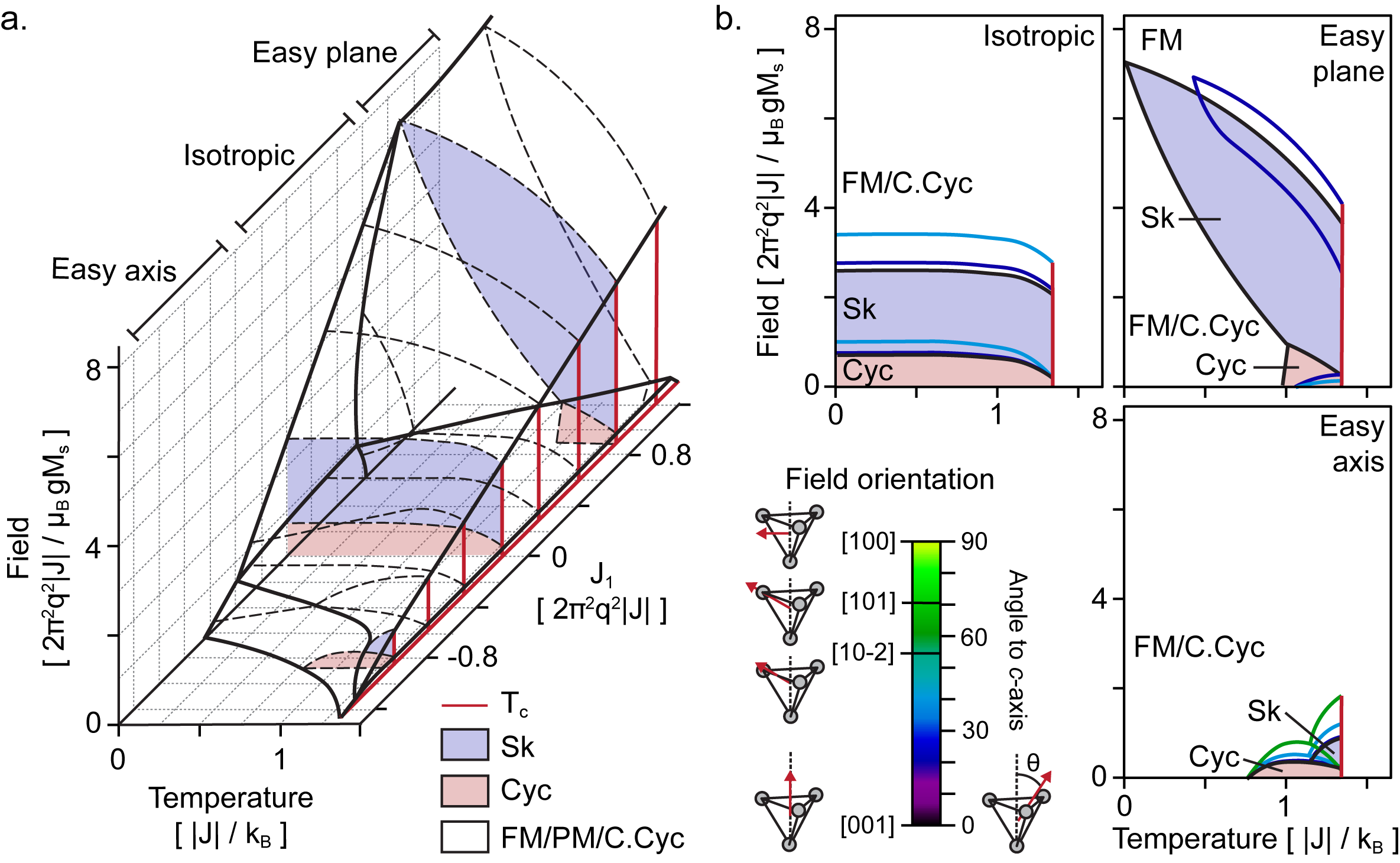}
\caption{\label{fig:ua} \textbf{a.} Variation in the normalized magnetic phase diagram with the uniaxial anisotropy parameter $J_1$, focusing on the cycloid and skyrmion phase transitions with applied field along the $c$-axis. Representative phase diagrams for the easy-axis, easy-plane, and isotropic cases ($J_1/2 \pi^2 q^2 |J| = -0.9, +0.9, 0$ respectively) are highlighted. Phase labels are defined in the caption to Figure \ref{fig:gms}. \textbf{b.} Variation in skyrmion phase boundaries with field orientation, for the easy-axis, easy-plane and isotropic cases. The color of the phase boundaries denote the angle of the field with respect to the $c$-axis.}
\end{figure*}

The perturbation data shown in the left panel of Figure \ref{fig:cc}b reveals that the low--temperature region of the phase diagram is largely invariant to changes in most of the $J$--coefficients in the Hamiltonian, save for changes in $|J|$ and $q$ which amount to rescaling the temperature and field axes. By far the most important correlation functions are those corresponding to uniaxial anisotropy ($\varphi_{1,3,6,10,15}$), which have a qualitatively similar impact on the phase boundaries, and higher--order in--plane anisotropy $\varphi_4$ (right panels). The uniaxial terms $\varphi_{1,3,6,10,15}$ shift the skyrmion/ferromagnet boundary, penalizing skyrmion formation in the easy--axis regime. The $\varphi_4$ anisotropy function is unique in that it only alters the phase boundaries left unaffected by the uniaxial terms, shifting the cycloid/skyrmion and canted cycloid/ferromagnet boundaries, but leaving the skyrmion/ferromagnet boundary unchanged. Fortuitously, the only anisotropy function included in the $J^{(0)}$ simulation is $\varphi_2$, which has no impact on the phase boundaries so that the $J^{(0)}$ results are equivalent to a zero--anisotropy regime where all helimagnetic phases of interest appear at all temperatures. Finally, the only impact of exchange and DMI terms beyond varying $q$ and $|J|$ is to alter the field at which a canted cycloid in the ($ab$)--plane transforms to a ferromagnet.

\subsection*{Skyrmion stability determined by uniaxial anisotropy}

At low temperature, the stability of skyrmions in this system is largely controlled by the presence of uniaxial anisotropy in the form of the $\varphi_{1,3,6,10,15}$ correlation functions, as well as high--order in--plane anisotropy in the form of $\varphi_4$. As uniaxial anisotropy is the most common form of anisotropy observed in uniaxial magnets, we now derive the impact of this term on the skyrmion region at all temperatures, using $\varphi_1$ as a proxy for all uniaxial anisotropy functions in the Hamiltonian. We note however that in rare cases where higher-order anisotropies ($\varphi_{2,...,6}$) are strong, more complex phase behavior is possible.

The evolution of the phase diagram with the coefficient of $\varphi_1$, $J_1$, as a function of field along the $c$--axis and temperature, is shown in Figure \ref{fig:ua}a. The phase diagram exhibits three broad regions corresponding to easy--axis, easy--plane, and isotropic scenarios. The skyrmion and cycloid regions are highlighted in blue and red respectively for representative slices in each region. We obtain this phase diagram using a linear extrapolation of the Helmholtz free energy $A$:
\[
A = A^{(0)} + \frac{\partial A}{\partial J_1} \Delta J_1 =  A^{(0)} + \langle \varphi_1 \rangle\Delta J_1 
\]
where $A^{(0)}$ is the Helmholtz free energy obtained for $J^{(0)}$. As this extrapolation is only applicable at constant temperature, we neglect any changes in $T_\text{c}$ due to changes in $J_1$. Similarly, as variation in $J_1$ has a negligible impact on $|J|$ and no impact on $q$, the normalization factors for field, temperature, and $J$ are taken to be constant. To check the validity of this extrapolation, we confirm that the extrapolated phase diagrams highlighted in the easy--axis and easy--plane regions of Figure \ref{fig:ua}a agree with Monte Carlo data for the same conditions.

The impact of uniaxial anisotropy on skyrmion stability is largely determined by the stabilization of the competing out--of--plane and in--plane ferromagnetic and canted cycloid phases. Easy--axis anisotropy favors the out--of--plane ferromagnet configuration and thus suppresses the skyrmion phase, and eventually the cycloid phase. Easy--plane anisotropy destabilizes the out--of--plane ferromagnet and thus enhances the skyrmion stability region, up to the point where the easy--plane--anisotropy is sufficient to stabilize canted cycloids and in--plane ferromagnetic configurations at zero--field, at which point the skyrmion region remains stable only at high $c$--axis fields,  consistent with phenomenological solutions\cite{Randeria2016,Utkan2016}. Elevated temperatures suppress the effect of anisotropy while preserving the qualitative trends, leading to a much wider range of anisotropy constants for which skyrmions and cycloids are stable than in the low--temperature limit. The profound impact of anisotropy constants on skyrmion stability at low temperature is reminiscent of recent reports of low--temperature skyrmion stabilization in Cu${}_2$OSeO${}_3$ where the role of anisotropy is to suppress the competing canted spin--wave phases.\cite{Chacon2018,Bannenberg2019}

Uniaxial anisotropy also varies in the range of field directions for which skyrmions may be observed, with easy--axis anisotropy favoring skyrmions over a wide range of field orientations and easy--plane allowing for skyrmions only for fields close to the $c$--axis. Figure \ref{fig:ua}b shows the evolution of the cycloid and skyrmion phase boundaries as a function of applied field direction. The color of the phase boundaries corresponds to the field angle, shown on the legend in terms of angle with respect to the $c$--axis and high--symmetry directions in the structure. In the isotropic case, the skyrmion region moves to slightly higher fields with increasing $\theta$, up to $\theta_{\text{max}} \approx 40 {}^\text{o}$, beyond which only cycloid, canted cycloid and ferromagnet phases are stable. In the easy--plane case, the skyrmion region quickly narrows with increasing $\theta$, disappearing above $\theta_{\text{max}} \approx 20 {}^\text{o}$. In the easy--axis case, the skyrmion stability region is much smaller in field--magnitude, but wider in field--angle, with a high--temperature skyrmion phase appearing up to $\theta_{\text{max}} \approx 70 {}^\text{o}$. These results, as well as the broad impact of uniaxial anisotropy, confirm the conclusions reached by a phenomenological analysis reported by Leonov and Kezsmarki\cite{Leonov2017}.

The formation of skyrmions in easy--axis systems over a wide range of field orientations is important in the context of resolving skyrmion formation experimentally. The easy--axis scenario leads to skyrmion formation for fields applied along the $[101]$ family of miller indices of the unit cell shown in Figure \ref{fig:structure}a. These directions correspond to the $\langle 001 \rangle$ axes of variants of the lacunar spinel distorted along directions equivalent to the $\langle 111 \rangle$ axis of the high--temperature $F\bar{4}3m$ phase. In a real material, where all symmetrically--equivalent variants of the $\langle 111 \rangle$ distortion are likely to be present, we would thus expect skyrmion phase boundaries following both the black and green curves, even if the field is applied parallel to the $c$--axis of one of the variants. This is precisely the scenario observed in GaV${}_4$S${}_8$\cite{Kezsmarki2015}, which we estimate corresponds to the strongly easy--axis $J_{1}/{2\pi^2 q^2 |J|} = -1.00$ slice of the phase diagram.

\subsection*{Predicted phase diagrams for V and Mo systems}

\begin{table}[t]
\begin{ruledtabular}
\begin{tabular}{l c c c c c}
	 				& $T_\text{c}$ (K) & $q$ 	& $J_1$ ($\mu$eV) 	& $J_{1}/{2\pi^2 q^2 |J|}$ 	\\ \hline
GaV${}_4$S${}_8$ 	& 13		& 1/26	& -25				& -1.00						\\
GaV${}_4$Se${}_8$ 	& 18		& 1/27	& 2					& 0.06 						\\
GaMo${}_4$S${}_8$ 	& 23		& 1/12	& -34				& -0.17						\\
GaMo${}_4$Se${}_8$ 	& 27		& 1/20	& 25				& 0.29						\\			
\end{tabular}
\end{ruledtabular}
\caption{\label{table:systems} Behavior of V and Mo--based lacunar spinels based on the phase diagram shown in Figure \ref{fig:ua}. $T_\text{c}$, $q$ and $J_1$ data for the V systems is based on experimental data from refs. \cite{Kezsmarki2015} and \cite{Bordacs2017}, while data for the Mo systems is based on our DFT calculations.}
\end{table}

Finally, we use the phase diagram shown in Figure \ref{fig:ua} to evaluate the magnetic phase diagrams of several V and Mo--based spinels. Our estimates for the value of the reduced anisotropy $J_{1}/{2\pi^2 q^2 |J|}$ for GaV${}_4$S${}_8$, GaV${}_4$Se${}_8$, GaMo${}_4$S${}_8$ and GaMo${}_4$Se${}_8$ are given in Table \ref{table:systems}, with phase diagrams given by field--temperature slices of Figure \ref{fig:ua}a at the given level of anisotropy. The magnetic phase diagrams of GaV${}_4$S${}_8$ and GaV${}_4$Se${}_8$ have been studied in--depth experimentally, and offer a direct  comparison to our results. We similarly rely on experimental data to obtain $T_\text{c}$, $q$ and $J_1$ for the V systems\cite{Kezsmarki2015, Bordacs2017}. Experimental data on GaMo${}_4$S${}_8$ and GaMo${}_4$Se${}_8$ is more sparse, but as these systems are well--represented by DFT, we use computational data to obtain estimates for the magnetic parameters. Given the small energy scale and critical importance of the anisotropy constants to phase behavior, we obtain the value of the anisotropy constants independently of other terms in the magnetic Hamiltonian, using data computed within the same unit cell so as to minimize numerical noise arising from k--point discretization error. Specifically, the energy of orienting a ferromagnetic configuration along various crystallographic directions, computed using the primitive cell of the structure, fully constrains $J_{1,...,6}$ independently of any exchange or DMI terms. Thus, we are able to refine the cluster expansion fit by first fitting $J_{1,...,6}$ to this high--accuracy anisotropy data, and then fitting all other terms to the full dataset. As a result, we are able to reduce the fitting error of the anisotropy energy to the level of $\mu$eV, as can be seen in Supplementary Figure 2\cite{SupplementaryMat}, sufficient to reliably resolve the anisotropy constants.

\section*{Discussion}

\subsection*{Broad applicability of the uniaxial skyrmion phase diagram}

A direct comparison of our predicted phase diagram to experimental results reported for GaV${}_4$S${}_8$ and GaV${}_4$Se${}_8$ lends credibility to our analysis. In both cases, the phase diagrams obtained using the reduced anisotropy values from Table \ref{table:systems} are in qualitative agreement with experiment. We quantitatively reproduce temperature behavior, but predict phase transitions at fields 1.5 to 2 times larger than those observed. The most likely source for this error is a deviation in $\mu_B g M_s$ away from our assumed value of $2 \mu_B$. We also neglect the impact of stray fields, which destabilize the cycloid and N\'eel skyrmion states seen here in favor of a ferromagnetic configuration with 180${}^\text{o}$ domain walls\cite{Bogdanov1994}. In the strong DMI regime relevant to lacunar spinels, this error does not qualitatively change the results. However, the impact of stray fields becomes pronounced for cycloidal systems with weak DMI, limiting the applicability of Figure \ref{fig:ua}a to systems well away from the long--wavelength cycloid stability bounds derived by Bogdanov and Hubert\cite{Bogdanov1994}. Furthermore, the agreement between the phase behavior we derive in Figure \ref{fig:ua} and that obtained phenomenologically for the same symmetry\cite{Randeria2016,Utkan2016,Leonov2017} provides an important consistency check for our atomistic model of the magnetic behavior of these skyrmion--host materials.

The insensitivity of the skyrmion region of the phase diagram to the values of DMI and exchange suggests that the phase behavior seen in Figure \ref{fig:ua}a may generalize to other uniaxial systems. We observe that high--moment N\'eel skyrmions form preferentially to canted cycloids when the magnetic field is orthogonal to the rotation axis of any cycloidal variant. This mechanism is independent of any specific interaction parameters and arises from the fact that canted cycloids only develop a moment parallel to their rotation axis as shown in Figure \ref{fig:structure}c. Thus, we speculate that a similar mechanism may lead to N\'eel skyrmion formation over wide field and temperature ranges in other systems where cycloid variants have rotation axes constrained to a single plane. Based on symmetry arguments alone, this behavior is most likely in strong--DMI, low--anisotropy systems whose point group is one of $6mm$ (C${}_{6\text{v}}$), $3m$ (C${}_{3\text{v}}$), $4mm$ (C${}_{4\text{v}}$), or $mm2$ (C${}_{2\text{v}}$) \cite{Kitchaev2018}. If we assume that a similar mechanism is applicable to the formation of canted helices and Bloch antiskyrmions, materials with $\bar{4}2m$ (D${}_{2\text{d}}$) and $\bar{4}$ (S${}_4$) point group symmetry may also exhibit this behavior. Several phenomenological analyses of skyrmion formation in C${}_{n\text{v}}$, D${}_{2\text{d}}$ and S${}_4$ crystals\cite{Randeria2016, Utkan2016, Leonov2017} report similar results, lending support to a broad applicability of these trends.

\subsection*{Parametrization of magnetic cluster expansion Hamiltonians}

An important implication of our results is that uncertainty quantification is essential to the construction of magnetic cluster expansion Hamiltonians. Conventional cluster expansion fitting techniques rely on the minimization of total error against DFT energies. This methodology works well when all interactions have similar energy scales, and when working with discrete degrees of freedom such as atomic configurations\cite{Zabaras2014}. However, this approach is not sufficient when relatively small energy terms, such as magnetocrystalline anisotropy explored here, play a decisive role in determining which phases form. By analyzing the sensitivity of the phase diagram to the values of the interaction coefficients through generalized Clausius-Clapeyron relations and linear free energy extrapolation based on $\partial A/\partial J_i = \langle \varphi_i \rangle$, one can identify terms in the cluster expansion which play an outsized role in determining phase behavior. In particular, one can use this approach to evaluate the importance of terms not included in the original Hamiltonian, either due to basis set truncation, or elimination during the fitting process. Once all important terms are known, the fitting procedure must be adjusted to ensure these terms are fitted accurately\cite{Mueller2009, Mueller2010}, which may increase the average error of the fit but nonetheless yields more qualitatively correct predictions of phase behavior.\\ 

\section*{Conclusion}

We have demonstrated that skyrmion stability across a wide range of fields and temperatures in the GaM${}_4$X${}_8$ lacunar spinels is a general consequence of the symmetry of the material and the fact that magnetocrystalline anisotropy energy is typically small. We reproduce the complex magnetic phase diagrams of these materials, including long--wavelength magnetic order, without relying on empirical parameters, and thus gain insight into the relationship between spin--orbit coupling and skyrmion formation. We find that magnetic cluster expansions parametrized using density functional theory data can accurately predict this magnetic phase diagram provided that the fitting procedure leads to a high--fidelity form for the anisotropy energy. More generally, reliable magnetic phase diagram prediction requires an evaluation of the impact of fitting error and uncertainty on phase stability. As we find that the magnetic phase behavior here is determined by simple, transferable mechanisms dictated by point--group symmetry, we speculate that our observations are likely broadly applicable to uniaxial systems with C${}_{n\text{v}}$, D${}_{2\text{d}}$ and S${}_4$ symmetry.

\begin{acknowledgments}
The research reported here was supported by the Materials Research Science and Engineering Center at UCSB (MRSEC NSF DMR 1720256) through IRG-1.
Computational resources for this project were provided by the National Energy Research Scientific Computing Center, a DOE Office of Science User Facility supported by the Office of Science of the U.S. Department of Energy under Contract No. DE-AC02-05CH11231, as well as the Center for Scientific Computing at UC Santa Barbara, which is supported by the National Science Foundation (NSF) Materials Research Science and Engineering Centers program through NSF DMR 1720256 and NSF CNS 1725797.
\end{acknowledgments}


\begin{thebibliography}{48}%
\makeatletter
\providecommand \@ifxundefined [1]{%
 \@ifx{#1\undefined}
}%
\providecommand \@ifnum [1]{%
 \ifnum #1\expandafter \@firstoftwo
 \else \expandafter \@secondoftwo
 \fi
}%
\providecommand \@ifx [1]{%
 \ifx #1\expandafter \@firstoftwo
 \else \expandafter \@secondoftwo
 \fi
}%
\providecommand \natexlab [1]{#1}%
\providecommand \enquote  [1]{``#1''}%
\providecommand \bibnamefont  [1]{#1}%
\providecommand \bibfnamefont [1]{#1}%
\providecommand \citenamefont [1]{#1}%
\providecommand \href@noop [0]{\@secondoftwo}%
\providecommand \href [0]{\begingroup \@sanitize@url \@href}%
\providecommand \@href[1]{\@@startlink{#1}\@@href}%
\providecommand \@@href[1]{\endgroup#1\@@endlink}%
\providecommand \@sanitize@url [0]{\catcode `\\12\catcode `\$12\catcode
  `\&12\catcode `\#12\catcode `\^12\catcode `\_12\catcode `\%12\relax}%
\providecommand \@@startlink[1]{}%
\providecommand \@@endlink[0]{}%
\providecommand \url  [0]{\begingroup\@sanitize@url \@url }%
\providecommand \@url [1]{\endgroup\@href {#1}{\urlprefix }}%
\providecommand \urlprefix  [0]{URL }%
\providecommand \Eprint [0]{\href }%
\providecommand \doibase [0]{http://dx.doi.org/}%
\providecommand \selectlanguage [0]{\@gobble}%
\providecommand \bibinfo  [0]{\@secondoftwo}%
\providecommand \bibfield  [0]{\@secondoftwo}%
\providecommand \translation [1]{[#1]}%
\providecommand \BibitemOpen [0]{}%
\providecommand \bibitemStop [0]{}%
\providecommand \bibitemNoStop [0]{.\EOS\space}%
\providecommand \EOS [0]{\spacefactor3000\relax}%
\providecommand \BibitemShut  [1]{\csname bibitem#1\endcsname}%
\let\auto@bib@innerbib\@empty
\bibitem [{\citenamefont {Bogdanov}\ and\ \citenamefont
  {Yablonskii}(1989)}]{Bogdanov1989}%
  \BibitemOpen
  \bibfield  {author} {\bibinfo {author} {\bibfnamefont {A.~N.}\ \bibnamefont
  {Bogdanov}}\ and\ \bibinfo {author} {\bibfnamefont {D.}~\bibnamefont
  {Yablonskii}},\ }\href@noop {} {\bibfield  {journal} {\bibinfo  {journal}
  {Sov. Phys. JETP}\ }\textbf {\bibinfo {volume} {68}},\ \bibinfo {pages} {101}
  (\bibinfo {year} {1989})}\BibitemShut {NoStop}%
\bibitem [{\citenamefont {Roessler}\ \emph {et~al.}(2006)\citenamefont
  {Roessler}, \citenamefont {Bogdanov},\ and\ \citenamefont
  {Pfleiderer}}]{Roessler2006}%
  \BibitemOpen
  \bibfield  {author} {\bibinfo {author} {\bibfnamefont {U.~K.}\ \bibnamefont
  {Roessler}}, \bibinfo {author} {\bibfnamefont {A.}~\bibnamefont {Bogdanov}},
  \ and\ \bibinfo {author} {\bibfnamefont {C.}~\bibnamefont {Pfleiderer}},\
  }\href {\doibase 10.1038/nature05056} {\bibfield  {journal} {\bibinfo
  {journal} {Nature}\ }\textbf {\bibinfo {volume} {442}},\ \bibinfo {pages}
  {797} (\bibinfo {year} {2006})}\BibitemShut {NoStop}%
\bibitem [{\citenamefont {M{\"u}hlbauer}\ \emph {et~al.}(2009)\citenamefont
  {M{\"u}hlbauer}, \citenamefont {Binz}, \citenamefont {Jonietz}, \citenamefont
  {Pfleiderer}, \citenamefont {Rosch}, \citenamefont {Neubauer}, \citenamefont
  {Georgii},\ and\ \citenamefont {B{\"o}ni}}]{Muhlbauer2009}%
  \BibitemOpen
  \bibfield  {author} {\bibinfo {author} {\bibfnamefont {S.}~\bibnamefont
  {M{\"u}hlbauer}}, \bibinfo {author} {\bibfnamefont {B.}~\bibnamefont {Binz}},
  \bibinfo {author} {\bibfnamefont {F.}~\bibnamefont {Jonietz}}, \bibinfo
  {author} {\bibfnamefont {C.}~\bibnamefont {Pfleiderer}}, \bibinfo {author}
  {\bibfnamefont {A.}~\bibnamefont {Rosch}}, \bibinfo {author} {\bibfnamefont
  {A.}~\bibnamefont {Neubauer}}, \bibinfo {author} {\bibfnamefont
  {R.}~\bibnamefont {Georgii}}, \ and\ \bibinfo {author} {\bibfnamefont
  {P.}~\bibnamefont {B{\"o}ni}},\ }\href@noop {} {\bibfield  {journal}
  {\bibinfo  {journal} {Science}\ }\textbf {\bibinfo {volume} {323}},\ \bibinfo
  {pages} {915} (\bibinfo {year} {2009})}\BibitemShut {NoStop}%
\bibitem [{\citenamefont {Jonietz}\ \emph {et~al.}(2010)\citenamefont
  {Jonietz}, \citenamefont {M{\"u}hlbauer}, \citenamefont {Pfleiderer},
  \citenamefont {Neubauer}, \citenamefont {M{\"u}nzer}, \citenamefont {Bauer},
  \citenamefont {Adams}, \citenamefont {Georgii}, \citenamefont {B{\"o}ni},
  \citenamefont {Duine}, \citenamefont {Everschor}, \citenamefont {Garst},\
  and\ \citenamefont {Rosch}}]{Jonietz12010}%
  \BibitemOpen
  \bibfield  {author} {\bibinfo {author} {\bibfnamefont {F.}~\bibnamefont
  {Jonietz}}, \bibinfo {author} {\bibfnamefont {S.}~\bibnamefont
  {M{\"u}hlbauer}}, \bibinfo {author} {\bibfnamefont {C.}~\bibnamefont
  {Pfleiderer}}, \bibinfo {author} {\bibfnamefont {A.}~\bibnamefont
  {Neubauer}}, \bibinfo {author} {\bibfnamefont {W.}~\bibnamefont
  {M{\"u}nzer}}, \bibinfo {author} {\bibfnamefont {A.}~\bibnamefont {Bauer}},
  \bibinfo {author} {\bibfnamefont {T.}~\bibnamefont {Adams}}, \bibinfo
  {author} {\bibfnamefont {R.}~\bibnamefont {Georgii}}, \bibinfo {author}
  {\bibfnamefont {P.}~\bibnamefont {B{\"o}ni}}, \bibinfo {author}
  {\bibfnamefont {R.~A.}\ \bibnamefont {Duine}}, \bibinfo {author}
  {\bibfnamefont {K.}~\bibnamefont {Everschor}}, \bibinfo {author}
  {\bibfnamefont {M.}~\bibnamefont {Garst}}, \ and\ \bibinfo {author}
  {\bibfnamefont {A.}~\bibnamefont {Rosch}},\ }\href {\doibase
  10.1126/science.1195709} {\bibfield  {journal} {\bibinfo  {journal}
  {Science}\ }\textbf {\bibinfo {volume} {330}},\ \bibinfo {pages} {1648}
  (\bibinfo {year} {2010})}\BibitemShut {NoStop}%
\bibitem [{\citenamefont {Sampaio}\ \emph {et~al.}(2013)\citenamefont
  {Sampaio}, \citenamefont {Cros}, \citenamefont {Rohart}, \citenamefont
  {Thiaville},\ and\ \citenamefont {Fert}}]{Sampaio2013}%
  \BibitemOpen
  \bibfield  {author} {\bibinfo {author} {\bibfnamefont {J.}~\bibnamefont
  {Sampaio}}, \bibinfo {author} {\bibfnamefont {V.}~\bibnamefont {Cros}},
  \bibinfo {author} {\bibfnamefont {S.}~\bibnamefont {Rohart}}, \bibinfo
  {author} {\bibfnamefont {A.}~\bibnamefont {Thiaville}}, \ and\ \bibinfo
  {author} {\bibfnamefont {A.}~\bibnamefont {Fert}},\ }\href {\doibase
  10.1038/nnano.2013.210} {\bibfield  {journal} {\bibinfo  {journal} {Nature
  nanotechnology}\ }\textbf {\bibinfo {volume} {8}},\ \bibinfo {pages} {839}
  (\bibinfo {year} {2013})}\BibitemShut {NoStop}%
\bibitem [{\citenamefont {Yu}\ \emph {et~al.}(2012)\citenamefont {Yu},
  \citenamefont {Kanazawa}, \citenamefont {Zhang}, \citenamefont {Nagai},
  \citenamefont {Hara}, \citenamefont {Kimoto}, \citenamefont {Matsui},
  \citenamefont {Onose},\ and\ \citenamefont {Tokura}}]{Yu2012}%
  \BibitemOpen
  \bibfield  {author} {\bibinfo {author} {\bibfnamefont {X.}~\bibnamefont
  {Yu}}, \bibinfo {author} {\bibfnamefont {N.}~\bibnamefont {Kanazawa}},
  \bibinfo {author} {\bibfnamefont {W.}~\bibnamefont {Zhang}}, \bibinfo
  {author} {\bibfnamefont {T.}~\bibnamefont {Nagai}}, \bibinfo {author}
  {\bibfnamefont {T.}~\bibnamefont {Hara}}, \bibinfo {author} {\bibfnamefont
  {K.}~\bibnamefont {Kimoto}}, \bibinfo {author} {\bibfnamefont
  {Y.}~\bibnamefont {Matsui}}, \bibinfo {author} {\bibfnamefont
  {Y.}~\bibnamefont {Onose}}, \ and\ \bibinfo {author} {\bibfnamefont
  {Y.}~\bibnamefont {Tokura}},\ }\href {\doibase 10.1038/ncomms1990} {\bibfield
   {journal} {\bibinfo  {journal} {Nature communications}\ }\textbf {\bibinfo
  {volume} {3}},\ \bibinfo {pages} {988} (\bibinfo {year} {2012})}\BibitemShut
  {NoStop}%
\bibitem [{\citenamefont {Seki}\ \emph {et~al.}(2012)\citenamefont {Seki},
  \citenamefont {Yu}, \citenamefont {Ishiwata},\ and\ \citenamefont
  {Tokura}}]{Seki2012}%
  \BibitemOpen
  \bibfield  {author} {\bibinfo {author} {\bibfnamefont {S.}~\bibnamefont
  {Seki}}, \bibinfo {author} {\bibfnamefont {X.~Z.}\ \bibnamefont {Yu}},
  \bibinfo {author} {\bibfnamefont {S.}~\bibnamefont {Ishiwata}}, \ and\
  \bibinfo {author} {\bibfnamefont {Y.}~\bibnamefont {Tokura}},\ }\href
  {\doibase 10.1126/science.1214143} {\bibfield  {journal} {\bibinfo  {journal}
  {Science}\ }\textbf {\bibinfo {volume} {336}},\ \bibinfo {pages} {198}
  (\bibinfo {year} {2012})}\BibitemShut {NoStop}%
\bibitem [{\citenamefont {Tokunaga}\ \emph {et~al.}(2015)\citenamefont
  {Tokunaga}, \citenamefont {Yu}, \citenamefont {White}, \citenamefont
  {R{\o}nnow}, \citenamefont {Morikawa}, \citenamefont {Taguchi},\ and\
  \citenamefont {Tokura}}]{Tokunaga2015}%
  \BibitemOpen
  \bibfield  {author} {\bibinfo {author} {\bibfnamefont {Y.}~\bibnamefont
  {Tokunaga}}, \bibinfo {author} {\bibfnamefont {X.}~\bibnamefont {Yu}},
  \bibinfo {author} {\bibfnamefont {J.}~\bibnamefont {White}}, \bibinfo
  {author} {\bibfnamefont {H.~M.}\ \bibnamefont {R{\o}nnow}}, \bibinfo {author}
  {\bibfnamefont {D.}~\bibnamefont {Morikawa}}, \bibinfo {author}
  {\bibfnamefont {Y.}~\bibnamefont {Taguchi}}, \ and\ \bibinfo {author}
  {\bibfnamefont {Y.}~\bibnamefont {Tokura}},\ }\href {\doibase
  10.1038/ncomms8638} {\bibfield  {journal} {\bibinfo  {journal} {Nature
  communications}\ }\textbf {\bibinfo {volume} {6}},\ \bibinfo {pages} {7638}
  (\bibinfo {year} {2015})}\BibitemShut {NoStop}%
\bibitem [{\citenamefont {Yu}\ \emph {et~al.}(2010)\citenamefont {Yu},
  \citenamefont {Onose}, \citenamefont {Kanazawa}, \citenamefont {Park},
  \citenamefont {Han}, \citenamefont {Matsui}, \citenamefont {Nagaosa},\ and\
  \citenamefont {Tokura}}]{Yu2010}%
  \BibitemOpen
  \bibfield  {author} {\bibinfo {author} {\bibfnamefont {X.}~\bibnamefont
  {Yu}}, \bibinfo {author} {\bibfnamefont {Y.}~\bibnamefont {Onose}}, \bibinfo
  {author} {\bibfnamefont {N.}~\bibnamefont {Kanazawa}}, \bibinfo {author}
  {\bibfnamefont {J.}~\bibnamefont {Park}}, \bibinfo {author} {\bibfnamefont
  {J.}~\bibnamefont {Han}}, \bibinfo {author} {\bibfnamefont {Y.}~\bibnamefont
  {Matsui}}, \bibinfo {author} {\bibfnamefont {N.}~\bibnamefont {Nagaosa}}, \
  and\ \bibinfo {author} {\bibfnamefont {Y.}~\bibnamefont {Tokura}},\ }\href
  {\doibase 10.1038/nature09124} {\bibfield  {journal} {\bibinfo  {journal}
  {Nature}\ }\textbf {\bibinfo {volume} {465}},\ \bibinfo {pages} {901}
  (\bibinfo {year} {2010})}\BibitemShut {NoStop}%
\bibitem [{\citenamefont {Fujima}\ \emph {et~al.}(2017)\citenamefont {Fujima},
  \citenamefont {Abe}, \citenamefont {Tokunaga},\ and\ \citenamefont
  {Arima}}]{Fujima2017}%
  \BibitemOpen
  \bibfield  {author} {\bibinfo {author} {\bibfnamefont {Y.}~\bibnamefont
  {Fujima}}, \bibinfo {author} {\bibfnamefont {N.}~\bibnamefont {Abe}},
  \bibinfo {author} {\bibfnamefont {Y.}~\bibnamefont {Tokunaga}}, \ and\
  \bibinfo {author} {\bibfnamefont {T.}~\bibnamefont {Arima}},\ }\href
  {\doibase 10.1103/PhysRevB.95.180410} {\bibfield  {journal} {\bibinfo
  {journal} {Phys. Rev. B}\ }\textbf {\bibinfo {volume} {95}},\ \bibinfo
  {pages} {180410} (\bibinfo {year} {2017})}\BibitemShut {NoStop}%
\bibitem [{\citenamefont {Bord{\'a}cs}\ \emph {et~al.}(2017)\citenamefont
  {Bord{\'a}cs}, \citenamefont {Butykai}, \citenamefont {Szigeti},
  \citenamefont {White}, \citenamefont {Cubitt}, \citenamefont {Leonov},
  \citenamefont {Widmann}, \citenamefont {Ehlers}, \citenamefont {von Nidda},
  \citenamefont {Tsurkan}, \citenamefont {Loidl},\ and\ \citenamefont
  {K{\'e}zsm{\'a}rki}}]{Bordacs2017}%
  \BibitemOpen
  \bibfield  {author} {\bibinfo {author} {\bibfnamefont {S.}~\bibnamefont
  {Bord{\'a}cs}}, \bibinfo {author} {\bibfnamefont {A.}~\bibnamefont
  {Butykai}}, \bibinfo {author} {\bibfnamefont {B.}~\bibnamefont {Szigeti}},
  \bibinfo {author} {\bibfnamefont {J.}~\bibnamefont {White}}, \bibinfo
  {author} {\bibfnamefont {R.}~\bibnamefont {Cubitt}}, \bibinfo {author}
  {\bibfnamefont {A.}~\bibnamefont {Leonov}}, \bibinfo {author} {\bibfnamefont
  {S.}~\bibnamefont {Widmann}}, \bibinfo {author} {\bibfnamefont
  {D.}~\bibnamefont {Ehlers}}, \bibinfo {author} {\bibfnamefont {H.-A.~K.}\
  \bibnamefont {von Nidda}}, \bibinfo {author} {\bibfnamefont {V.}~\bibnamefont
  {Tsurkan}}, \bibinfo {author} {\bibfnamefont {A.}~\bibnamefont {Loidl}}, \
  and\ \bibinfo {author} {\bibfnamefont {I.}~\bibnamefont
  {K{\'e}zsm{\'a}rki}},\ }\href {\doibase 10.1038/s41598-017-07996-x}
  {\bibfield  {journal} {\bibinfo  {journal} {Scientific Reports}\ }\textbf
  {\bibinfo {volume} {7}},\ \bibinfo {pages} {7584} (\bibinfo {year}
  {2017})}\BibitemShut {NoStop}%
\bibitem [{\citenamefont {Nayak}\ \emph {et~al.}(2017)\citenamefont {Nayak},
  \citenamefont {Kumar}, \citenamefont {Ma}, \citenamefont {Werner},
  \citenamefont {Pippel}, \citenamefont {Sahoo}, \citenamefont {Damay},
  \citenamefont {R{\"o}{\ss}ler}, \citenamefont {Felser},\ and\ \citenamefont
  {Parkin}}]{Nayak2017}%
  \BibitemOpen
  \bibfield  {author} {\bibinfo {author} {\bibfnamefont {A.~K.}\ \bibnamefont
  {Nayak}}, \bibinfo {author} {\bibfnamefont {V.}~\bibnamefont {Kumar}},
  \bibinfo {author} {\bibfnamefont {T.}~\bibnamefont {Ma}}, \bibinfo {author}
  {\bibfnamefont {P.}~\bibnamefont {Werner}}, \bibinfo {author} {\bibfnamefont
  {E.}~\bibnamefont {Pippel}}, \bibinfo {author} {\bibfnamefont
  {R.}~\bibnamefont {Sahoo}}, \bibinfo {author} {\bibfnamefont
  {F.}~\bibnamefont {Damay}}, \bibinfo {author} {\bibfnamefont {U.~K.}\
  \bibnamefont {R{\"o}{\ss}ler}}, \bibinfo {author} {\bibfnamefont
  {C.}~\bibnamefont {Felser}}, \ and\ \bibinfo {author} {\bibfnamefont {S.~S.}\
  \bibnamefont {Parkin}},\ }\href {\doibase 10.1038/nature23466} {\bibfield
  {journal} {\bibinfo  {journal} {Nature}\ }\textbf {\bibinfo {volume} {548}},\
  \bibinfo {pages} {561} (\bibinfo {year} {2017})}\BibitemShut {NoStop}%
\bibitem [{\citenamefont {K{\'e}zsm{\'a}rki}\ \emph {et~al.}(2015)\citenamefont
  {K{\'e}zsm{\'a}rki}, \citenamefont {Bord{\'a}cs}, \citenamefont {Milde},
  \citenamefont {Neuber}, \citenamefont {Eng}, \citenamefont {White},
  \citenamefont {R{\o}nnow}, \citenamefont {Dewhurst}, \citenamefont
  {Mochizuki}, \citenamefont {Yanai}, \citenamefont {Nakamura}, \citenamefont
  {Ehlers}, \citenamefont {Tsurkan},\ and\ \citenamefont
  {Loidl}}]{Kezsmarki2015}%
  \BibitemOpen
  \bibfield  {author} {\bibinfo {author} {\bibfnamefont {I.}~\bibnamefont
  {K{\'e}zsm{\'a}rki}}, \bibinfo {author} {\bibfnamefont {S.}~\bibnamefont
  {Bord{\'a}cs}}, \bibinfo {author} {\bibfnamefont {P.}~\bibnamefont {Milde}},
  \bibinfo {author} {\bibfnamefont {E.}~\bibnamefont {Neuber}}, \bibinfo
  {author} {\bibfnamefont {L.~M.}\ \bibnamefont {Eng}}, \bibinfo {author}
  {\bibfnamefont {J.~S.}\ \bibnamefont {White}}, \bibinfo {author}
  {\bibfnamefont {H.~M.}\ \bibnamefont {R{\o}nnow}}, \bibinfo {author}
  {\bibfnamefont {C.~D.}\ \bibnamefont {Dewhurst}}, \bibinfo {author}
  {\bibfnamefont {M.}~\bibnamefont {Mochizuki}}, \bibinfo {author}
  {\bibfnamefont {K.}~\bibnamefont {Yanai}}, \bibinfo {author} {\bibfnamefont
  {H.}~\bibnamefont {Nakamura}}, \bibinfo {author} {\bibfnamefont
  {D.}~\bibnamefont {Ehlers}}, \bibinfo {author} {\bibfnamefont
  {V.}~\bibnamefont {Tsurkan}}, \ and\ \bibinfo {author} {\bibfnamefont
  {A.}~\bibnamefont {Loidl}},\ }\href {\doibase 10.1038/nmat4402} {\bibfield
  {journal} {\bibinfo  {journal} {Nature Materials}\ }\textbf {\bibinfo
  {volume} {14}},\ \bibinfo {pages} {1116} (\bibinfo {year}
  {2015})}\BibitemShut {NoStop}%
\bibitem [{\citenamefont {Zhang}\ \emph {et~al.}(2019)\citenamefont {Zhang},
  \citenamefont {Chen}, \citenamefont {Barone}, \citenamefont {Yamauchi},
  \citenamefont {Dong},\ and\ \citenamefont {Picozzi}}]{Zhang2019}%
  \BibitemOpen
  \bibfield  {author} {\bibinfo {author} {\bibfnamefont {H.-M.}\ \bibnamefont
  {Zhang}}, \bibinfo {author} {\bibfnamefont {J.}~\bibnamefont {Chen}},
  \bibinfo {author} {\bibfnamefont {P.}~\bibnamefont {Barone}}, \bibinfo
  {author} {\bibfnamefont {K.}~\bibnamefont {Yamauchi}}, \bibinfo {author}
  {\bibfnamefont {S.}~\bibnamefont {Dong}}, \ and\ \bibinfo {author}
  {\bibfnamefont {S.}~\bibnamefont {Picozzi}},\ }\href {\doibase
  10.1103/PhysRevB.99.214427} {\bibfield  {journal} {\bibinfo  {journal} {Phys.
  Rev. B}\ }\textbf {\bibinfo {volume} {99}},\ \bibinfo {pages} {214427}
  (\bibinfo {year} {2019})}\BibitemShut {NoStop}%
\bibitem [{\citenamefont {Schueller}\ \emph {et~al.}(2019)\citenamefont
  {Schueller}, \citenamefont {Zuo}, \citenamefont {Bocarsly}, \citenamefont
  {Kitchaev}, \citenamefont {Wilson},\ and\ \citenamefont
  {Seshadri}}]{Schueller2019}%
  \BibitemOpen
  \bibfield  {author} {\bibinfo {author} {\bibfnamefont {E.~C.}\ \bibnamefont
  {Schueller}}, \bibinfo {author} {\bibfnamefont {J.~L.}\ \bibnamefont {Zuo}},
  \bibinfo {author} {\bibfnamefont {J.~D.}\ \bibnamefont {Bocarsly}}, \bibinfo
  {author} {\bibfnamefont {D.~A.}\ \bibnamefont {Kitchaev}}, \bibinfo {author}
  {\bibfnamefont {S.~D.}\ \bibnamefont {Wilson}}, \ and\ \bibinfo {author}
  {\bibfnamefont {R.}~\bibnamefont {Seshadri}},\ }\href {\doibase
  10.1103/PhysRevB.100.045131} {\bibfield  {journal} {\bibinfo  {journal}
  {Phys. Rev. B}\ }\textbf {\bibinfo {volume} {100}},\ \bibinfo {pages}
  {045131} (\bibinfo {year} {2019})}\BibitemShut {NoStop}%
\bibitem [{\citenamefont {Streltsov}\ and\ \citenamefont
  {Khomskii}(2016)}]{Streltsov10491}%
  \BibitemOpen
  \bibfield  {author} {\bibinfo {author} {\bibfnamefont {S.~V.}\ \bibnamefont
  {Streltsov}}\ and\ \bibinfo {author} {\bibfnamefont {D.~I.}\ \bibnamefont
  {Khomskii}},\ }\href {\doibase 10.1073/pnas.1606367113} {\bibfield  {journal}
  {\bibinfo  {journal} {Proceedings of the National Academy of Sciences}\
  }\textbf {\bibinfo {volume} {113}},\ \bibinfo {pages} {10491} (\bibinfo
  {year} {2016})}\BibitemShut {NoStop}%
\bibitem [{\citenamefont {Kim}\ \emph {et~al.}(2014)\citenamefont {Kim},
  \citenamefont {Im}, \citenamefont {Han},\ and\ \citenamefont
  {Jin}}]{Kim2014}%
  \BibitemOpen
  \bibfield  {author} {\bibinfo {author} {\bibfnamefont {H.-S.}\ \bibnamefont
  {Kim}}, \bibinfo {author} {\bibfnamefont {J.}~\bibnamefont {Im}}, \bibinfo
  {author} {\bibfnamefont {M.~J.}\ \bibnamefont {Han}}, \ and\ \bibinfo
  {author} {\bibfnamefont {H.}~\bibnamefont {Jin}},\ }\href {\doibase
  10.1038/ncomms4988 (2014)} {\bibfield  {journal} {\bibinfo  {journal} {Nature
  communications}\ }\textbf {\bibinfo {volume} {5}},\ \bibinfo {pages} {3988}
  (\bibinfo {year} {2014})}\BibitemShut {NoStop}%
\bibitem [{\citenamefont {Sanchez}\ \emph {et~al.}(1984)\citenamefont
  {Sanchez}, \citenamefont {Ducastelle},\ and\ \citenamefont
  {Gratias}}]{Sanchez1984}%
  \BibitemOpen
  \bibfield  {author} {\bibinfo {author} {\bibfnamefont {J.}~\bibnamefont
  {Sanchez}}, \bibinfo {author} {\bibfnamefont {F.}~\bibnamefont {Ducastelle}},
  \ and\ \bibinfo {author} {\bibfnamefont {D.}~\bibnamefont {Gratias}},\ }\href
  {\doibase 10.1016/0378-4371(84)90096-7} {\bibfield  {journal} {\bibinfo
  {journal} {Physica A: Statistical Mechanics and its Applications}\ }\textbf
  {\bibinfo {volume} {128}},\ \bibinfo {pages} {334 } (\bibinfo {year}
  {1984})}\BibitemShut {NoStop}%
\bibitem [{\citenamefont {van~de Walle}\ and\ \citenamefont
  {Ceder}(2002)}]{vandeWalle2002}%
  \BibitemOpen
  \bibfield  {author} {\bibinfo {author} {\bibfnamefont {A.}~\bibnamefont
  {van~de Walle}}\ and\ \bibinfo {author} {\bibfnamefont {G.}~\bibnamefont
  {Ceder}},\ }\href {\doibase 10.1361/105497102770331596} {\bibfield  {journal}
  {\bibinfo  {journal} {Journal of Phase Equilibria}\ }\textbf {\bibinfo
  {volume} {23}},\ \bibinfo {pages} {348} (\bibinfo {year} {2002})}\BibitemShut
  {NoStop}%
\bibitem [{\citenamefont {Drautz}\ and\ \citenamefont
  {F\"ahnle}(2004)}]{Drautz2004}%
  \BibitemOpen
  \bibfield  {author} {\bibinfo {author} {\bibfnamefont {R.}~\bibnamefont
  {Drautz}}\ and\ \bibinfo {author} {\bibfnamefont {M.}~\bibnamefont
  {F\"ahnle}},\ }\href {\doibase 10.1103/PhysRevB.69.104404} {\bibfield
  {journal} {\bibinfo  {journal} {Phys. Rev. B}\ }\textbf {\bibinfo {volume}
  {69}},\ \bibinfo {pages} {104404} (\bibinfo {year} {2004})}\BibitemShut
  {NoStop}%
\bibitem [{\citenamefont {Mueller}\ and\ \citenamefont
  {Ceder}(2006)}]{Mueller2006}%
  \BibitemOpen
  \bibfield  {author} {\bibinfo {author} {\bibfnamefont {T.}~\bibnamefont
  {Mueller}}\ and\ \bibinfo {author} {\bibfnamefont {G.}~\bibnamefont
  {Ceder}},\ }\href {\doibase 10.1103/PhysRevB.74.134104} {\bibfield  {journal}
  {\bibinfo  {journal} {Phys. Rev. B}\ }\textbf {\bibinfo {volume} {74}},\
  \bibinfo {pages} {134104} (\bibinfo {year} {2006})}\BibitemShut {NoStop}%
\bibitem [{\citenamefont {Janson}\ \emph {et~al.}(2014)\citenamefont {Janson},
  \citenamefont {Rousochatzakis}, \citenamefont {Tsirlin}, \citenamefont
  {Belesi}, \citenamefont {Leonov}, \citenamefont {Roßler}, \citenamefont {Van
  Den~Brink},\ and\ \citenamefont {Rosner}}]{Janson2016}%
  \BibitemOpen
  \bibfield  {author} {\bibinfo {author} {\bibfnamefont {O.}~\bibnamefont
  {Janson}}, \bibinfo {author} {\bibfnamefont {I.}~\bibnamefont
  {Rousochatzakis}}, \bibinfo {author} {\bibfnamefont {A.~A.}\ \bibnamefont
  {Tsirlin}}, \bibinfo {author} {\bibfnamefont {M.}~\bibnamefont {Belesi}},
  \bibinfo {author} {\bibfnamefont {A.~A.}\ \bibnamefont {Leonov}}, \bibinfo
  {author} {\bibfnamefont {U.~K.}\ \bibnamefont {Roßler}}, \bibinfo {author}
  {\bibfnamefont {J.}~\bibnamefont {Van Den~Brink}}, \ and\ \bibinfo {author}
  {\bibfnamefont {H.}~\bibnamefont {Rosner}},\ }\href@noop {} {\bibfield
  {journal} {\bibinfo  {journal} {Nature communications}\ }\textbf {\bibinfo
  {volume} {5}},\ \bibinfo {pages} {5376} (\bibinfo {year} {2014})}\BibitemShut
  {NoStop}%
\bibitem [{\citenamefont {Van~der Ven}\ \emph {et~al.}(2018)\citenamefont
  {Van~der Ven}, \citenamefont {Thomas}, \citenamefont {Puchala},\ and\
  \citenamefont {Natarajan}}]{VdV2018}%
  \BibitemOpen
  \bibfield  {author} {\bibinfo {author} {\bibfnamefont {A.}~\bibnamefont
  {Van~der Ven}}, \bibinfo {author} {\bibfnamefont {J.}~\bibnamefont {Thomas}},
  \bibinfo {author} {\bibfnamefont {B.}~\bibnamefont {Puchala}}, \ and\
  \bibinfo {author} {\bibfnamefont {A.}~\bibnamefont {Natarajan}},\ }\href
  {\doibase 10.1146/annurev-matsci-070317-124443} {\bibfield  {journal}
  {\bibinfo  {journal} {Annual Review of Materials Research}\ }\textbf
  {\bibinfo {volume} {48}},\ \bibinfo {pages} {27} (\bibinfo {year}
  {2018})}\BibitemShut {NoStop}%
\bibitem [{\citenamefont {Bogdanov}\ and\ \citenamefont
  {Hubert}(1994)}]{Bogdanov1994}%
  \BibitemOpen
  \bibfield  {author} {\bibinfo {author} {\bibfnamefont {A.}~\bibnamefont
  {Bogdanov}}\ and\ \bibinfo {author} {\bibfnamefont {A.}~\bibnamefont
  {Hubert}},\ }\href {\doibase 10.1016/0304-8853(94)90046-9} {\bibfield
  {journal} {\bibinfo  {journal} {Journal of Magnetism and Magnetic Materials}\
  }\textbf {\bibinfo {volume} {138}},\ \bibinfo {pages} {255 } (\bibinfo {year}
  {1994})}\BibitemShut {NoStop}%
\bibitem [{\citenamefont {Thomas}\ and\ \citenamefont {Van~der
  Ven}(2017)}]{Thomas2017}%
  \BibitemOpen
  \bibfield  {author} {\bibinfo {author} {\bibfnamefont {J.~C.}\ \bibnamefont
  {Thomas}}\ and\ \bibinfo {author} {\bibfnamefont {A.}~\bibnamefont {Van~der
  Ven}},\ }\href@noop {} {\bibfield  {journal} {\bibinfo  {journal} {Journal of
  the Mechanics and Physics of Solids}\ }\textbf {\bibinfo {volume} {107}},\
  \bibinfo {pages} {76} (\bibinfo {year} {2017})}\BibitemShut {NoStop}%
\bibitem [{\citenamefont {Thomas}\ \emph {et~al.}(2018)\citenamefont {Thomas},
  \citenamefont {Bechtel},\ and\ \citenamefont {Van~der Ven}}]{Thomas2018}%
  \BibitemOpen
  \bibfield  {author} {\bibinfo {author} {\bibfnamefont {J.~C.}\ \bibnamefont
  {Thomas}}, \bibinfo {author} {\bibfnamefont {J.~S.}\ \bibnamefont {Bechtel}},
  \ and\ \bibinfo {author} {\bibfnamefont {A.}~\bibnamefont {Van~der Ven}},\
  }\href {\doibase 10.1103/PhysRevB.98.094105} {\bibfield  {journal} {\bibinfo
  {journal} {Phys. Rev. B}\ }\textbf {\bibinfo {volume} {98}},\ \bibinfo
  {pages} {094105} (\bibinfo {year} {2018})}\BibitemShut {NoStop}%
\bibitem [{\citenamefont {Singer}\ \emph {et~al.}(2011)\citenamefont {Singer},
  \citenamefont {Dietermann},\ and\ \citenamefont {F\"ahnle}}]{Singer2011}%
  \BibitemOpen
  \bibfield  {author} {\bibinfo {author} {\bibfnamefont {R.}~\bibnamefont
  {Singer}}, \bibinfo {author} {\bibfnamefont {F.}~\bibnamefont {Dietermann}},
  \ and\ \bibinfo {author} {\bibfnamefont {M.}~\bibnamefont {F\"ahnle}},\
  }\href {\doibase 10.1103/PhysRevLett.107.017204} {\bibfield  {journal}
  {\bibinfo  {journal} {Phys. Rev. Lett.}\ }\textbf {\bibinfo {volume} {107}},\
  \bibinfo {pages} {017204} (\bibinfo {year} {2011})}\BibitemShut {NoStop}%
\bibitem [{\citenamefont {Lam}\ \emph {et~al.}(2015)\citenamefont {Lam},
  \citenamefont {Pitrou},\ and\ \citenamefont {Seibert}}]{Numba2015}%
  \BibitemOpen
  \bibfield  {author} {\bibinfo {author} {\bibfnamefont {S.~K.}\ \bibnamefont
  {Lam}}, \bibinfo {author} {\bibfnamefont {A.}~\bibnamefont {Pitrou}}, \ and\
  \bibinfo {author} {\bibfnamefont {S.}~\bibnamefont {Seibert}},\ }in\ \href
  {\doibase 10.1145/2833157.2833162} {\emph {\bibinfo {booktitle} {Proceedings
  of the Second Workshop on the LLVM Compiler Infrastructure in HPC}}},\
  \bibinfo {series and number} {LLVM '15}\ (\bibinfo  {publisher} {ACM},\
  \bibinfo {year} {2015})\ pp.\ \bibinfo {pages} {7:1--7:6}\BibitemShut
  {NoStop}%
\bibitem [{\citenamefont {Ong}\ \emph {et~al.}(2013)\citenamefont {Ong},
  \citenamefont {Richards}, \citenamefont {Jain}, \citenamefont {Hautier},
  \citenamefont {Kocher}, \citenamefont {Cholia}, \citenamefont {Gunter},
  \citenamefont {Chevrier}, \citenamefont {Persson},\ and\ \citenamefont
  {Ceder}}]{Ong2013}%
  \BibitemOpen
  \bibfield  {author} {\bibinfo {author} {\bibfnamefont {S.~P.}\ \bibnamefont
  {Ong}}, \bibinfo {author} {\bibfnamefont {W.~D.}\ \bibnamefont {Richards}},
  \bibinfo {author} {\bibfnamefont {A.}~\bibnamefont {Jain}}, \bibinfo {author}
  {\bibfnamefont {G.}~\bibnamefont {Hautier}}, \bibinfo {author} {\bibfnamefont
  {M.}~\bibnamefont {Kocher}}, \bibinfo {author} {\bibfnamefont
  {S.}~\bibnamefont {Cholia}}, \bibinfo {author} {\bibfnamefont
  {D.}~\bibnamefont {Gunter}}, \bibinfo {author} {\bibfnamefont {V.~L.}\
  \bibnamefont {Chevrier}}, \bibinfo {author} {\bibfnamefont {K.~A.}\
  \bibnamefont {Persson}}, \ and\ \bibinfo {author} {\bibfnamefont
  {G.}~\bibnamefont {Ceder}},\ }\href {\doibase
  10.1016/j.commatsci.2012.10.028} {\bibfield  {journal} {\bibinfo  {journal}
  {Comp. Mater. Sci.}\ }\textbf {\bibinfo {volume} {68}},\ \bibinfo {pages}
  {314 } (\bibinfo {year} {2013})}\BibitemShut {NoStop}%
\bibitem [{Sup()}]{SupplementaryMat}%
  \BibitemOpen
  \href@noop {} {\bibinfo  {journal} {See Supplementary Materials at [URL will
  be inserted by publisher] for derivation details and complete form of the
  cluster expansion Hamiltonian, derivation of the field and temperature
  normalization constants, map of the skyrmion topological index across field
  and temperature, and quantification of error in parametrizing DFT data}\
  }\BibitemShut {NoStop}%
\bibitem [{\citenamefont {Hart}\ \emph {et~al.}(2005)\citenamefont {Hart},
  \citenamefont {Blum}, \citenamefont {Walorski},\ and\ \citenamefont
  {Zunger}}]{Hart2005}%
  \BibitemOpen
\bibfield  {journal} {  }\bibfield  {author} {\bibinfo {author} {\bibfnamefont
  {G.~L.}\ \bibnamefont {Hart}}, \bibinfo {author} {\bibfnamefont
  {V.}~\bibnamefont {Blum}}, \bibinfo {author} {\bibfnamefont {M.~J.}\
  \bibnamefont {Walorski}}, \ and\ \bibinfo {author} {\bibfnamefont
  {A.}~\bibnamefont {Zunger}},\ }\href {\doibase 10.1038/nmat1374} {\bibfield
  {journal} {\bibinfo  {journal} {Nature materials}\ }\textbf {\bibinfo
  {volume} {4}},\ \bibinfo {pages} {391} (\bibinfo {year} {2005})}\BibitemShut
  {NoStop}%
\bibitem [{\citenamefont {Wang}\ \emph {et~al.}(2019)\citenamefont {Wang},
  \citenamefont {Hammerschmidt}, \citenamefont {Rogal},\ and\ \citenamefont
  {Drautz}}]{Drautz2019}%
  \BibitemOpen
  \bibfield  {author} {\bibinfo {author} {\bibfnamefont {N.}~\bibnamefont
  {Wang}}, \bibinfo {author} {\bibfnamefont {T.}~\bibnamefont {Hammerschmidt}},
  \bibinfo {author} {\bibfnamefont {J.}~\bibnamefont {Rogal}}, \ and\ \bibinfo
  {author} {\bibfnamefont {R.}~\bibnamefont {Drautz}},\ }\href {\doibase
  10.1103/PhysRevB.99.094402} {\bibfield  {journal} {\bibinfo  {journal} {Phys.
  Rev. B}\ }\textbf {\bibinfo {volume} {99}},\ \bibinfo {pages} {094402}
  (\bibinfo {year} {2019})}\BibitemShut {NoStop}%
\bibitem [{\citenamefont {Hoffman}\ and\ \citenamefont
  {Gelman}(2014)}]{Hoffman2014}%
  \BibitemOpen
  \bibfield  {author} {\bibinfo {author} {\bibfnamefont {M.~D.}\ \bibnamefont
  {Hoffman}}\ and\ \bibinfo {author} {\bibfnamefont {A.}~\bibnamefont
  {Gelman}},\ }\href@noop {} {\bibfield  {journal} {\bibinfo  {journal}
  {Journal of Machine Learning Research}\ }\textbf {\bibinfo {volume} {15}},\
  \bibinfo {pages} {1593} (\bibinfo {year} {2014})}\BibitemShut {NoStop}%
\bibitem [{\citenamefont {Betancourt}(2017)}]{Betancourt2017}%
  \BibitemOpen
  \bibfield  {author} {\bibinfo {author} {\bibfnamefont {M.}~\bibnamefont
  {Betancourt}},\ }\href@noop {} {\bibfield  {journal} {\bibinfo  {journal}
  {arXiv preprint arXiv:1701.02434}\ } (\bibinfo {year} {2017})}\BibitemShut
  {NoStop}%
\bibitem [{\citenamefont {Kresse}\ and\ \citenamefont
  {Furthmüller}(1996)}]{Kresse1996a}%
  \BibitemOpen
  \bibfield  {author} {\bibinfo {author} {\bibfnamefont {G.}~\bibnamefont
  {Kresse}}\ and\ \bibinfo {author} {\bibfnamefont {J.}~\bibnamefont
  {Furthmüller}},\ }\href {\doibase 10.1016/0927-0256(96)00008-0} {\bibfield
  {journal} {\bibinfo  {journal} {Computational Materials Science}\ }\textbf
  {\bibinfo {volume} {6}},\ \bibinfo {pages} {15 } (\bibinfo {year}
  {1996})}\BibitemShut {NoStop}%
\bibitem [{\citenamefont {Kresse}\ and\ \citenamefont
  {Joubert}(1999)}]{Kresse1999}%
  \BibitemOpen
  \bibfield  {author} {\bibinfo {author} {\bibfnamefont {G.}~\bibnamefont
  {Kresse}}\ and\ \bibinfo {author} {\bibfnamefont {D.}~\bibnamefont
  {Joubert}},\ }\href {\doibase 10.1103/PhysRevB.59.1758} {\bibfield  {journal}
  {\bibinfo  {journal} {Phys. Rev. B}\ }\textbf {\bibinfo {volume} {59}},\
  \bibinfo {pages} {1758} (\bibinfo {year} {1999})}\BibitemShut {NoStop}%
\bibitem [{\citenamefont {Perdew}\ \emph {et~al.}(1996)\citenamefont {Perdew},
  \citenamefont {Burke},\ and\ \citenamefont {Ernzerhof}}]{PBE1996}%
  \BibitemOpen
  \bibfield  {author} {\bibinfo {author} {\bibfnamefont {J.~P.}\ \bibnamefont
  {Perdew}}, \bibinfo {author} {\bibfnamefont {K.}~\bibnamefont {Burke}}, \
  and\ \bibinfo {author} {\bibfnamefont {M.}~\bibnamefont {Ernzerhof}},\ }\href
  {\doibase 10.1103/physrevlett.77.3865} {\bibfield  {journal} {\bibinfo
  {journal} {Physical review letters}\ }\textbf {\bibinfo {volume} {77}},\
  \bibinfo {pages} {3865} (\bibinfo {year} {1996})}\BibitemShut {NoStop}%
\bibitem [{\citenamefont {Brazovskii}(1975)}]{Brazovskii1975}%
  \BibitemOpen
  \bibfield  {author} {\bibinfo {author} {\bibfnamefont {S.~A.}\ \bibnamefont
  {Brazovskii}},\ }\href@noop {} {\bibfield  {journal} {\bibinfo  {journal}
  {Sov. Phys. JETP}\ }\textbf {\bibinfo {volume} {41}},\ \bibinfo {pages} {85}
  (\bibinfo {year} {1975})}\BibitemShut {NoStop}%
\bibitem [{\citenamefont {Janoschek}\ \emph {et~al.}(2013)\citenamefont
  {Janoschek}, \citenamefont {Garst}, \citenamefont {Bauer}, \citenamefont
  {Krautscheid}, \citenamefont {Georgii}, \citenamefont {B\"oni},\ and\
  \citenamefont {Pfleiderer}}]{Janoschek2013}%
  \BibitemOpen
  \bibfield  {author} {\bibinfo {author} {\bibfnamefont {M.}~\bibnamefont
  {Janoschek}}, \bibinfo {author} {\bibfnamefont {M.}~\bibnamefont {Garst}},
  \bibinfo {author} {\bibfnamefont {A.}~\bibnamefont {Bauer}}, \bibinfo
  {author} {\bibfnamefont {P.}~\bibnamefont {Krautscheid}}, \bibinfo {author}
  {\bibfnamefont {R.}~\bibnamefont {Georgii}}, \bibinfo {author} {\bibfnamefont
  {P.}~\bibnamefont {B\"oni}}, \ and\ \bibinfo {author} {\bibfnamefont
  {C.}~\bibnamefont {Pfleiderer}},\ }\href {\doibase
  10.1103/PhysRevB.87.134407} {\bibfield  {journal} {\bibinfo  {journal} {Phys.
  Rev. B}\ }\textbf {\bibinfo {volume} {87}},\ \bibinfo {pages} {134407}
  (\bibinfo {year} {2013})}\BibitemShut {NoStop}%
\bibitem [{\citenamefont {Rowland}\ \emph {et~al.}(2016)\citenamefont
  {Rowland}, \citenamefont {Banerjee},\ and\ \citenamefont
  {Randeria}}]{Randeria2016}%
  \BibitemOpen
  \bibfield  {author} {\bibinfo {author} {\bibfnamefont {J.}~\bibnamefont
  {Rowland}}, \bibinfo {author} {\bibfnamefont {S.}~\bibnamefont {Banerjee}}, \
  and\ \bibinfo {author} {\bibfnamefont {M.}~\bibnamefont {Randeria}},\ }\href
  {\doibase 10.1103/PhysRevB.93.020404} {\bibfield  {journal} {\bibinfo
  {journal} {Phys. Rev. B}\ }\textbf {\bibinfo {volume} {93}},\ \bibinfo
  {pages} {020404} (\bibinfo {year} {2016})}\BibitemShut {NoStop}%
\bibitem [{\citenamefont {G\"ung\"ord\"u}\ \emph {et~al.}(2016)\citenamefont
  {G\"ung\"ord\"u}, \citenamefont {Nepal}, \citenamefont {Tretiakov},
  \citenamefont {Belashchenko},\ and\ \citenamefont {Kovalev}}]{Utkan2016}%
  \BibitemOpen
  \bibfield  {author} {\bibinfo {author} {\bibfnamefont {U.}~\bibnamefont
  {G\"ung\"ord\"u}}, \bibinfo {author} {\bibfnamefont {R.}~\bibnamefont
  {Nepal}}, \bibinfo {author} {\bibfnamefont {O.~A.}\ \bibnamefont
  {Tretiakov}}, \bibinfo {author} {\bibfnamefont {K.}~\bibnamefont
  {Belashchenko}}, \ and\ \bibinfo {author} {\bibfnamefont {A.~A.}\
  \bibnamefont {Kovalev}},\ }\href {\doibase 10.1103/PhysRevB.93.064428}
  {\bibfield  {journal} {\bibinfo  {journal} {Phys. Rev. B}\ }\textbf {\bibinfo
  {volume} {93}},\ \bibinfo {pages} {064428} (\bibinfo {year}
  {2016})}\BibitemShut {NoStop}%
\bibitem [{\citenamefont {Chacon}\ \emph {et~al.}(2018)\citenamefont {Chacon},
  \citenamefont {Heinen}, \citenamefont {Halder}, \citenamefont {Bauer},
  \citenamefont {Simeth}, \citenamefont {M{\"u}hlbauer}, \citenamefont
  {Berger}, \citenamefont {Garst}, \citenamefont {Rosch},\ and\ \citenamefont
  {Pfleiderer}}]{Chacon2018}%
  \BibitemOpen
  \bibfield  {author} {\bibinfo {author} {\bibfnamefont {A.}~\bibnamefont
  {Chacon}}, \bibinfo {author} {\bibfnamefont {L.}~\bibnamefont {Heinen}},
  \bibinfo {author} {\bibfnamefont {M.}~\bibnamefont {Halder}}, \bibinfo
  {author} {\bibfnamefont {A.}~\bibnamefont {Bauer}}, \bibinfo {author}
  {\bibfnamefont {W.}~\bibnamefont {Simeth}}, \bibinfo {author} {\bibfnamefont
  {S.}~\bibnamefont {M{\"u}hlbauer}}, \bibinfo {author} {\bibfnamefont
  {H.}~\bibnamefont {Berger}}, \bibinfo {author} {\bibfnamefont
  {M.}~\bibnamefont {Garst}}, \bibinfo {author} {\bibfnamefont
  {A.}~\bibnamefont {Rosch}}, \ and\ \bibinfo {author} {\bibfnamefont
  {C.}~\bibnamefont {Pfleiderer}},\ }\href {\doibase 10.1038/s41567-018-0184-y}
  {\bibfield  {journal} {\bibinfo  {journal} {Nature Physics}\ }\textbf
  {\bibinfo {volume} {14}},\ \bibinfo {pages} {936} (\bibinfo {year}
  {2018})}\BibitemShut {NoStop}%
\bibitem [{\citenamefont {Bannenberg}\ \emph {et~al.}(2019)\citenamefont
  {Bannenberg}, \citenamefont {Wilhelm}, \citenamefont {Cubitt}, \citenamefont
  {Labh}, \citenamefont {Schmidt}, \citenamefont {Leli{\`e}vre-Berna},
  \citenamefont {Pappas}, \citenamefont {Mostovoy},\ and\ \citenamefont
  {Leonov}}]{Bannenberg2019}%
  \BibitemOpen
  \bibfield  {author} {\bibinfo {author} {\bibfnamefont {L.~J.}\ \bibnamefont
  {Bannenberg}}, \bibinfo {author} {\bibfnamefont {H.}~\bibnamefont {Wilhelm}},
  \bibinfo {author} {\bibfnamefont {R.}~\bibnamefont {Cubitt}}, \bibinfo
  {author} {\bibfnamefont {A.}~\bibnamefont {Labh}}, \bibinfo {author}
  {\bibfnamefont {M.~P.}\ \bibnamefont {Schmidt}}, \bibinfo {author}
  {\bibfnamefont {E.}~\bibnamefont {Leli{\`e}vre-Berna}}, \bibinfo {author}
  {\bibfnamefont {C.}~\bibnamefont {Pappas}}, \bibinfo {author} {\bibfnamefont
  {M.}~\bibnamefont {Mostovoy}}, \ and\ \bibinfo {author} {\bibfnamefont
  {A.~O.}\ \bibnamefont {Leonov}},\ }\href {\doibase 10.1038/s41535-019-0150-7}
  {\bibfield  {journal} {\bibinfo  {journal} {npj Quantum Materials}\ }\textbf
  {\bibinfo {volume} {4}},\ \bibinfo {pages} {11} (\bibinfo {year}
  {2019})}\BibitemShut {NoStop}%
\bibitem [{\citenamefont {Leonov}\ and\ \citenamefont
  {K\'ezsm\'arki}(2017)}]{Leonov2017}%
  \BibitemOpen
  \bibfield  {author} {\bibinfo {author} {\bibfnamefont {A.~O.}\ \bibnamefont
  {Leonov}}\ and\ \bibinfo {author} {\bibfnamefont {I.}~\bibnamefont
  {K\'ezsm\'arki}},\ }\href {\doibase 10.1103/PhysRevB.96.214413} {\bibfield
  {journal} {\bibinfo  {journal} {Phys. Rev. B}\ }\textbf {\bibinfo {volume}
  {96}},\ \bibinfo {pages} {214413} (\bibinfo {year} {2017})}\BibitemShut
  {NoStop}%
\bibitem [{\citenamefont {Kitchaev}\ \emph {et~al.}(2018)\citenamefont
  {Kitchaev}, \citenamefont {Beyerlein},\ and\ \citenamefont {Van~der
  Ven}}]{Kitchaev2018}%
  \BibitemOpen
  \bibfield  {author} {\bibinfo {author} {\bibfnamefont {D.~A.}\ \bibnamefont
  {Kitchaev}}, \bibinfo {author} {\bibfnamefont {I.~J.}\ \bibnamefont
  {Beyerlein}}, \ and\ \bibinfo {author} {\bibfnamefont {A.}~\bibnamefont
  {Van~der Ven}},\ }\href {\doibase 10.1103/PhysRevB.98.214414} {\bibfield
  {journal} {\bibinfo  {journal} {Phys. Rev. B}\ }\textbf {\bibinfo {volume}
  {98}},\ \bibinfo {pages} {214414} (\bibinfo {year} {2018})}\BibitemShut
  {NoStop}%
\bibitem [{\citenamefont {Kristensen}\ and\ \citenamefont
  {Zabaras}(2014)}]{Zabaras2014}%
  \BibitemOpen
  \bibfield  {author} {\bibinfo {author} {\bibfnamefont {J.}~\bibnamefont
  {Kristensen}}\ and\ \bibinfo {author} {\bibfnamefont {N.~J.}\ \bibnamefont
  {Zabaras}},\ }\href {\doibase 10.1016/j.cpc.2014.07.013} {\bibfield
  {journal} {\bibinfo  {journal} {Computer Physics Communications}\ }\textbf
  {\bibinfo {volume} {185}},\ \bibinfo {pages} {2885 } (\bibinfo {year}
  {2014})}\BibitemShut {NoStop}%
\bibitem [{\citenamefont {Mueller}\ and\ \citenamefont
  {Ceder}(2009)}]{Mueller2009}%
  \BibitemOpen
  \bibfield  {author} {\bibinfo {author} {\bibfnamefont {T.}~\bibnamefont
  {Mueller}}\ and\ \bibinfo {author} {\bibfnamefont {G.}~\bibnamefont
  {Ceder}},\ }\href {\doibase 10.1103/PhysRevB.80.024103} {\bibfield  {journal}
  {\bibinfo  {journal} {Phys. Rev. B}\ }\textbf {\bibinfo {volume} {80}},\
  \bibinfo {pages} {024103} (\bibinfo {year} {2009})}\BibitemShut {NoStop}%
\bibitem [{\citenamefont {Mueller}\ and\ \citenamefont
  {Ceder}(2010)}]{Mueller2010}%
  \BibitemOpen
  \bibfield  {author} {\bibinfo {author} {\bibfnamefont {T.}~\bibnamefont
  {Mueller}}\ and\ \bibinfo {author} {\bibfnamefont {G.}~\bibnamefont
  {Ceder}},\ }\href {\doibase 10.1103/PhysRevB.82.184107} {\bibfield  {journal}
  {\bibinfo  {journal} {Phys. Rev. B}\ }\textbf {\bibinfo {volume} {82}},\
  \bibinfo {pages} {184107} (\bibinfo {year} {2010})}\BibitemShut {NoStop}%
\end{thebibliography}
%

\end{document}